\documentclass{emulateapj}
\usepackage{psfrag}
\usepackage{graphicx,epsfig,psfig}
\usepackage[normalem]{ulem}
\usepackage[dvipsnames]{xcolor}
\usepackage[]{inputenc,amssymb}
\usepackage{amsmath}
\usepackage{color}
\usepackage{epsfig}
\usepackage{textcomp}
\usepackage{xcolor,cancel}

\definecolor{webgreen}{rgb}{0,.5,0}
\definecolor{webbrown}{rgb}{.6,0,0}
\usepackage[pdfpagelabels]{hyperref}
\hypersetup{%
   colorlinks=true,hyperfootnotes=false,%
   breaklinks=true,%
   plainpages=false, bookmarksnumbered, bookmarksopen=true,
   bookmarksopenlevel=1,%
   urlcolor=webbrown, linkcolor=RoyalBlue, citecolor=webgreen,
   }
\usepackage[toc,page]{appendix}

\defcitealias{sha12}{S12}

\newcommand{\myemail}{deovrat@physics.iisc.ernet.in}
\newcommand{\msun}{\; {\rm M}_\odot}
\newcommand{\vect}[1]{\boldsymbol{#1}}
\newcommand{\lan}{\langle}
\newcommand{\ran}{\rangle}
\newcommand{\tr}{$t_{\rm cool}/t_{\rm ff}$}

\shorttitle{Cold gas and AGN jet feedback in cluster cores}
\shortauthors{D. Prasad, P. Sharma, A. Babul}

\begin{document}

\title{Cool core cycles: Cold gas and AGN jet feedback in cluster cores}

\author{Deovrat Prasad$^1$}\email{\myemail} \author{Prateek Sharma$^1$}\email{prateek@physics.iisc.ernet.in}
\affil{$^1$Joint Astronomy Program and Department of Physics, Indian Institute of Science,
    Bangalore, India 560012}
\author{Arif Babul$^2$}\email{babul@uvic.ca}
\affil{$^2$Department of Physics and Astronomy, University of Victoria, Victoria, BC V8P 1A1, Canada}

\begin{abstract}
Using high-resolution 3-D and 2-D (axisymmetric) hydrodynamic simulations in spherical geometry, we study the evolution of cool cluster cores heated by 
feedback-driven bipolar active galactic nuclei (AGN) jets. Condensation of cold gas, and
the consequent enhanced accretion, is required for AGN feedback to balance radiative cooling with reasonable efficiencies, and to match the observed cool
core properties. 
A feedback efficiency (mechanical luminosity $\approx \epsilon \dot{M}_{\rm acc} c^2$; where $\dot{M}_{\rm acc}$ is the mass accretion rate at 1 kpc) 
as small as $6 \times 10^{-5}$ is sufficient to reduce the cooling/accretion rate by $\sim 10$ compared to a pure cooling flow in clusters (with M$_{200} 
\lesssim 7 \times 10^{14}$ M$_\odot$). 
This value is {much} smaller compared to the ones considered earlier, and is consistent with the jet efficiency and the fact that only a 
small fraction of gas at 1 kpc is accreted on to the supermassive black hole (SMBH). {The feedback efficiency in earlier works was so high that the cluster 
core reached equilibrium in a hot state without much precipitation, unlike what is observed in cool-core clusters.} 
We find hysteresis
cycles in all our simulations with cold mode feedback:
{\em condensation} of cold
gas when the ratio of the cooling-time to the free-fall time ($t_{\rm cool}/t_{\rm ff}$) is $\lesssim 10$ leads to a sudden enhancement in the accretion rate; 
a large accretion rate causes strong jets and {\em overheating} of the hot ICM such that $t_{\rm cool}/t_{\rm ff} > 10$; further condensation of cold gas is 
suppressed and the accretion rate falls, leading to slow cooling of the core and condensation of cold gas, restarting the cycle. Therefore, there is a spread in
core properties, such as the jet power, accretion rate, for the same value of core entropy or \tr. A fewer number of cycles are observed for higher efficiencies
 and for lower mass halos because the core is overheated to a longer cooling time. The 3-D simulations show the formation
of a few-kpc scale, rotationally-supported, massive ($\sim 10^{11} \msun$) cold gas torus. Since the torus gas is not accreted on to the SMBH, it is largely 
decoupled from the feedback cycle. The radially dominant cold gas ($T<5\times 10^4$ K; $|v_r|>|v_\phi|$) consists of fast cold gas uplifted by AGN 
jets and freely-infalling cold gas condensing out of the core. The radially dominant cold gas extends out to 25 kpc {for the fiducial run (halo mass $7\times 
10^{14} \msun$ and feedback efficiency $6 \times 10^{-5}$)},  
with the average mass inflow 
rate dominating the outflow rate by a  factor of $\approx 2$.  We compare our simulation results with recent observations.
\end{abstract}


\section{Introduction}

Majority of baryons in galaxy clusters are in the form of a hot plasma known as the 
intracluster medium (ICM). In absence of cooling and heating, the ICM is expected to follow self-similar profiles for density, 
temperature, etc., irrespective of the halo mass (\citealt{kai86,kai91}; see also the review by \citealt{voi05}). 
However, self-similarity is not observed in either groups or clusters (e.g., \citealt{pon99,bal99,bab02}).
Moreover, the core cooling times in about a third of clusters is shorter than 1 Gyr, much shorter than their
age ($\sim$Hubble time; e.g., \citealt{cav09,pra09}). Thus, we expect cooling to shape the distribution of baryons in these cool-core clusters.

The existence of cool cores with short cooling times in a good fraction of galaxy clusters is a long-standing puzzle. {According to the classical cooling flow model, 
cluster cores} with such short cooling times were expected to cool catastrophically and to fuel star formation at a rate of $100-1000 \msun$ yr$^{-1}$ 
(e.g., \citealt{fab94,lew00}). 
However, cooling, dropout, and star formation at these high rates are never seen in cluster cores  (e.g., \citealt{edg01,pet03,ode08}). This means 
that some source(s) of heating is(are) able to replenish the core cooling losses, thereby preventing runaway cooling and star-formation.

While there are potential heat sources, such as the kinetic energy of in-falling galaxies and 
sub-halos (e.g., \citealt{dek08}), thermal conduction from the hotter outskirts (e.g., \citealt{voi04,voi11}), a {\em globally stable} mechanism, which 
increases rapidly with an increasing hot gas density in the core, is required to prevent catastrophic cooling. Observations of several cool-core 
clusters by {\em Chandra} and {\em XMM-Newton} have uncovered AGN-jet-driven X-ray cavities, whose mechanical power is enough to balance 
radiative cooling in the core (e.g., \citealt{boh02,bir04,mcn07}). The AGN jets are powered by the accretion of the cooling ICM on to the supermassive black hole
(SMBH) at the center of the dominant cluster galaxy. Thus, more cooling/accretion leads to an enhanced jet power and ICM heating, closing a 
 feedback loop that prevents runaway cooling in the core.

AGN feedback has been long-suspected to play a role in self-regulating the ICM (e.g., \citealt{bin95,cio01,sok01,bab02,mcc08}), but a clear picture has emerged 
only recently. While AGN feedback should provide feedback heating in cluster cores (as it is enhanced with ICM cooling), it is not obvious if, 
for reasonable parameters, 
AGN heating can keep pace with cooling that increases rapidly with an increasing core density.  Moreover, the dense core gas is expected to be highly 
susceptible to  fragmentation, leading to the formation of a multiphase medium consisting of cold dense clouds condensing from the hot diffuse
 ICM itself.  \citet{piz05} suggest that AGN outbursts that result in the heating of the cluster cores are due to the infall and accretion of these cold clumps.

The importance of cold gas precipitation/feedback has also been been highlighted
by several recent observations.
The fact that there is some multiphase-cooling/star-formation, albeit at a much smaller rate than predicted by the cooling flow 
estimate (\citealt{sok01}), ties well with the idea of a small fraction of the thermally unstable core gas cooling to the stable atomic and molecular temperatures.
A lot of this cold gas is expected to form stars, but some should be accreted on to the central SMBH.
Reservoirs of atomic (e.g., \citealt{cra99,mcd11,wer14}) and molecular gas (e.g., \citealt{don00,edg01,sal06,rus14,osu15}), both extended 
and centrally concentrated, and ongoing star formation (e.g., \citealt{bil08,hic10,mcd11b}) are observed in a lot of cool-core clusters.
Additionally, powerful radio jets/bubbles observed in most cool-core clusters (\citealt{cav08,mit09}) can be interpreted as a signature of kinetic feedback due
to cold gas accretion on to the SMBH.

Since cool cores are in rough global thermal balance (i.e.,  the cooling rate minus the heating rate is smaller than just the radiative cooling rate), the existence of 
cold gas in cluster cores can be understood as a consequence of local thermal instability in a weakly stratified atmosphere (\citealt{mcc12,sha12,sin15}). The 
idealized simulations, which impose global thermal equilibrium in the ICM, show that the nonlinear evolution of local thermal instability leads to 
in-situ condensation of cold gas only if the ratio of the cooling time and the free-fall time (\tr) is $\lesssim 10$ (\citealt{sha12}).

This model has the attractive feature that once the local thermal instability sets in and the cold gas begins to condenses out of the dense ICM,  it typically falls 
freely toward the center.   Some of the infalling cold gas has sufficiently low angular momentum to be accreted by the SMBH, resulting in the cold phase mass 
accretion rate onto SMBHs that can  exceed the  hot/Bondi accretion rate by a factor $\sim 10-100$ (\citealt{gas13,sha12}). This enhanced accretion rate in the cold phase can explain both the global thermal balance in cluster cores and the general lack of massive cooling flows in almost all cool-core clusters whereas, the hot-mode (Bondi) accretion rate appears inadequate by orders of magnitude  (e.g., \citealt{mcn11}). 

In detail, the precipitation of the cold gas, followed by a  sudden increase in the accretion rate onto the SMBHs, leads to an increase in jet/cavity power and 
(slight) overheating of the core.  The core expands and as the ratio \tr~rises above the threshold value of \tr=10, the gas is no longer prone to 
condensation.   The accretion rate drops, as does the jet power.   The core cools slowly and the whole cycle starts again when \tr $\lesssim10$. The 
frequency of heating/cooling cycles depend on jet efficiency and the halo mass. These features of the cold feedback model are verified in our 
numerical simulations. 

In fact, the  simple criterion of \tr$\lesssim 10$ for the onset of local thermal instability is expected to be generic -- applicable not only to the intracluster 
medium (ICM) but also the intragroup medium (IGrM)  and the circumgalactic medium (CGM) of all galaxies, including the Milky way (\citealt{sha12b,voi15}).  
This, in turn, has far-reaching implications for providing a common framework for understanding the the breaking of self-similarity in the properties of hot gas 
across the hierarchy, from galaxies to groups to clusters,  the presence of multi-phase gas in group and clusters cores, and the detection of cold gas in 
galaxies at distances of $\sim100$ kpc (e.g., \citealt{werk14}).  In fact, recent more realistic AGN jet feedback simulations show that cold gas condensation 
begins when the  \tr$\lesssim 10$ condition is met, and two distinct cold gas structures emerge: extended cold filaments which go out 10s of kpc; and a 
few-kpc rotationally-supported cold torus (\citealt{gas12,li14,li14b}).  This dichotomy in cold gas distribution is also seen in observations (e.g., \citealt{mcd11}). 

Now that the theoretical models are satisfactorily able to describe the basic state of the ICM in cool cluster cores, and since observations of cold gas and 
jets/cavities are rapidly accumulating, it is ripe to make detailed comparisons between observations and numerical simulations. We also aim to investigate
the similarities and differences in cold gas and jet/bubble properties as a function of the halo mass and feedback efficiency.

{In this paper, we focus on cool-core clusters} and have carried out 3-D and 2-D (axisymmetric) simulations of the interaction of feedback-driven AGN jets 
with the ICM over cosmological 
timescales, varying the halo mass and the feedback efficiency. The 3-D simulations, which should correspond more closely to reality, 
show the formation of a cold, massive, angular-momentum-supported  torus, as seen in previous works (\citealt{gas12,li14,li14b}). This massive cold torus is 
decoupled from the AGN feedback cycle, which is governed by the low angular momentum, radially-dominant ($|v_r|>|v_\phi|$, $v_r/v_\phi$ is the 
radial/azimuthal component of the velocity) in-falling cold gas. Angular-momentum-supported
gas is absent in 2-D simulations because of axisymmetry and the absence of rotation in the initial state (stochastic angular momentum can be generated in 
3-D because of $\partial/\partial \phi$ terms in the angular momentum equation). However, 2-D simulations are useful for two reasons: 
first, they show similar behavior to 3-D 
simulations, if we only consider the radially-dominant ($|v_r|>|v_\phi|$) cold gas; second, they are much cheaper to run for long timescales, and thus 
are useful to do parameter scans in halo mass and accretion efficiency. 

Compared to previous works (\citealt{gas12,li14,li14b}), we have carried out simulations with smaller feedback jet efficiencies.
We find that a feedback efficiency as low as $6 \times 10^{-5}$ (ratio of the input jet power and $\dot{M}_{\rm acc} c^2$, 
where $\dot{M}_{\rm acc}$ is the accretion rate measured at 1 kpc) is sufficient to reduce the mass accretion/cooling
rate by a factor of about 10 compared to the cooling flow value in groups and clusters. Such a low feedback efficiency fits in nicely with the observations 
which suggest that only a small fraction ($\sim 0.01$) of the available gas is accreted by the SMBH (e.g., \citealt{loe01}), and with the 
estimate of jet efficiency ($\sim 0.001-0.01$) with respect to the SMBH accretion rate (e.g., \citealt{ben09}). 
Moreover, jets in SMBHs are observed predominantly when the accretion rate is $\lesssim 0.01$ the Eddington 
value (e.g., \citealt{nar95,mer03}); i.e., $\lesssim 0.22 \msun$ yr$^{-1}$ for a $10^9 \msun$ SMBH. The expected mass accretion rate on to the SMBH in our simulations 
($\sim$0.01 times $\dot{M}_{\rm acc}$ in Table \ref{table}) satisfies this constraint.

We have analyzed the velocity-radius distribution of the cold gas in our simulations to compare with recent {\em ALMA} and {\em Herschel} observations of
cold gas structure and kinematics in galaxy/cluster cores (e.g., \citealt{mcn14,dav14,wer14}). Our simulations help in interpreting
observations of cold gas outflows and inflows at scales $\gtrsim 10$ kpc, and the rotationally-supported cold torus at scales $\lesssim 5$ kpc. In our
simulations, the fast ($\gtrsim 500$ km s$^{-1}$) atomic/molecular outflows are uplifted by the outgoing AGN jet. The slower ($\lesssim 300$ km s$^{-1}$) 
infall of cold gas is due to condensation in the dense core. The cold gas in the rotationally supported torus is at the local circular 
velocity ($\sim 200$ km s$^{-1}$).

Our paper is organized as follows. In section \ref{sec:num} we present the numerical setup, in particular our implementation of mass and kinetic energy
injection due to AGN jets. Section \ref{sec:res} presents the key results from our 3-D and 2-D simulations, a comparison of 3-D vs. 2-D, and the impact
of parameters such as feedback efficiency and halo mass on our results. In section \ref{sec:dis} we discuss our results and compare with previous simulations
and observations, and we conclude with a brief summary in section \ref{sec:con}.

\section{Numerical setup \& governing equations}
\label{sec:num}

We modify the ZEUS-MP code, a widely-used finite-difference MHD code  (\citealt{hay06}), to simulate cooling and AGN feedback cycles in galaxy clusters.
We solve the standard hydrodynamic equations using spherical $(r,~\theta,~\phi)$ coordinates, with cooling, external gravity, and mass and momentum source 
terms due to AGN feedback:
\begin{eqnarray}
\label{eq:mass}
&& \frac{\partial{\rho}}{\partial t} + {\bf \nabla \cdot} ( \rho \vect{ v} ) = S_{\rho}, \\
\label{eq:momentum}
&& \frac{\partial \rho \vect{v}}{\partial t} + {\bf \nabla}.(\rho \vect{v}\otimes \vect{v}) = - {\bf \nabla} p -\rho{\bf \nabla} \Phi + S_{\rho} v_{\rm jet} \vect{\hat{r}}, \\
\label{eq:int_energy}
&& e \frac{d}{dt} \ln(p/\rho^\gamma) = -n_e n_i \Lambda(T),
\end{eqnarray}
where $\rho$ is the mass density, $\vect{v}$ is the fluid velocity, $p=(\gamma-1)e$ is the pressure ($e$ is the internal energy density and $\gamma=5/3$ is 
the adiabatic index), $\Lambda(T)$ 
is the temperature-dependent cooling function, $n_e~(n_i)$ is the electron (ion) number density given by $\rho/[\mu_{e(i)} m_p]$ ($\mu_e=1.18$ and $\mu_i=1.3$
are the mean molecular weights per electron and per ion, respectively, for the ICM with a third solar metallicity). For the cooling function, we use a fit proposed in \citealt{sha10} (their Eq. 12 and solid line in their Fig. 1) with a stable phase at $10^4$ K.

In addition to the terms shown in Eqs. \ref{eq:mass}-\ref{eq:int_energy}, the code uses the standard explicit artificial viscosity, and has implicit 
diffusion associated with the numerical scheme (\citealt{sto92}). In addition to the standard non-linear viscosity, we use the linear viscosity, as recommended
by \citet{hay06} for strong shocks (see their Appendix B3.2). 

We use a fixed external NFW gravitational potential $\Phi(\vect{r})$ due to the dark matter halo (\citealt{nav96});
\begin{equation}
\label{eq:NFW}
\Phi(r) = - \frac{GM_{200}}{r}\frac{\ln(1+c_{200} r/r_{200})}{[\ln(1+c_{200}) - c_{200}/(1+c_{200})]} ,
\end{equation}
where $M_{200}$ ($r_{200}$) is the characteristic halo mass (radius) and $c_{200}\equiv r_{200}/r_s$ is the concentration parameter; the dark matter density
within $r_{200}$ is 200 times the critical density of the universe and $r_s$ is the scale radius. In this paper we focus on cluster and massive cluster runs 
with $M_{200}=7\times 10^{14} \msun$ and $1.8 \times 10^{15} \msun$, respectively, and adopt $c_{200}=4.7$ for all models. 

We include the source terms $S_\rho$ for mass and $S_\rho v_{\rm jet} \vect{\hat{r}}$ for the 
radial momentum to drive AGN jets ($v_{\rm jet}$ is the velocity which the jet matter is put in).\footnote{We have also carried out narrow-jet 
simulations with momentum injection in the vertical $[\vect{\hat{z}}]$ direction, but do not find much difference 
from our runs with momentum injection in the radial $[\vect{\hat{r}};{\rm ~see~Eq.~\ref{eq:momentum}}]$ direction.}
These source terms and the cooling term 
(in Eq. \ref{eq:int_energy}) are applied in 
an operator-split fashion. The mass and momentum source terms are approximated forward in time and centered in space. The cooling term is applied 
using a semi-implicit method described in Eq. 7 of \citet{mcc12}.

{Our simulations do not include physical processes like star formation and supernova  feedback. Star formation may deplete some of the cold gas available
in the cores (see \citealt{li15}), but this is unlikely to change our results for a realistic model of star formation. 
Supernova feedback is energetically subdominant compared to AGN
feedback, and cannot realistically suppress cluster cooling flows (e.g., \citealt{sar06}). We only include the most relevant physical processes, namely 
cooling and AGN jet feedback, in our present simulations.} 

\subsection{Jet Implementation}
Jets are implemented in the active domain by adding mass and momentum source terms as shown in Eqs. \ref{eq:mass} \& \ref{eq:momentum}. The source 
terms are negligible outside a small biconical region centered at the origin around $\theta=0,~\pi$, mimicking mass and momentum injection by fast bipolar 
AGN jets.

The density source term is implemented as
$$
S_{\rho} (r,\theta)= {\cal N} \dot{M}_{\rm jet}\psi (r,\theta),
$$
where $\dot{M}_{\rm jet}$ is the {\em single-jet} mass loading rate,
\begin{eqnarray}
\nonumber
\psi (r,\theta) &=& \left [ 2 +\tanh \left ( \frac{\theta_{\rm jet}-\theta}{\sigma_{\rm \theta}} \right ) + \tanh \left ( \frac{\theta_{\rm jet}+\theta -\pi}{\sigma_{\rm \theta}} \right )  \right ] \\
\label{eq:psi}
&& \times \left [1+\tanh \left (\frac{r_{\rm jet}-r}{\sigma_r} \right) \right] \times \frac{1}{4} 
\end{eqnarray}
describes the spatial distribution of the source term which falls smoothly to zero outside the small biconical jet region of radius $r_{\rm jet}$ and half-opening 
angle $\theta_{\rm jet}$. We smooth the jet source terms in space 
because the Kelvin-Helmholtz instability is known to be suppressed due to numerical diffusion in a fast flow if the shear layer is unresolved (e.g., \citealt{rob10}).
The normalization factor
$$
{\cal N} = \frac{3}{2 \pi r_{\rm jet}^3 (1- \cos\theta_{\rm jet})}
$$
ensures that the total mass added due to jets per unit time is $2 \dot{M}_{\rm jet}$. All our
simulations use the following jet parameters: $\sigma_r = 0.05$ kpc, $\theta_{\rm jet}=\pi/6$, and $\sigma_\theta =0.05$.  The jet source region with an opening 
angle of 30 degrees may sound large but we get similar results with narrower jets.  Also, the fast jet extends well beyond the source region and is much narrower 
(c.f. third panel in Figure \ref{fig:3-D_fiducial}). 
The jet radius $r_{\rm jet}$ is scaled with the halo mass; i.e., 
$$
r_{\rm jet}=2~{\rm kpc}~\left ( \frac{M_{200}}{7\times 10^{14} \msun} \right )^{1/3}.
$$

The jet mass-loading rate is calculated from the current mass accretion rate ($\dot{M}_{\rm acc}$) evaluated at the inner radial boundary 
such that the increase in the jet kinetic energy is a fixed fraction 
of the energy released via accretion; i.e., 
\begin{equation}
\label{eq:fb}
\dot{M}_{\rm jet}v_{ \rm jet}^2 = \epsilon \dot{M}_{\rm acc}c^2.
\end{equation}
We choose the jet velocity $v_{\rm jet}=3 \times 10^4$ km s$^{-1}$ ($0.1 c$; $c$ is the speed of light); such fast velocities are seen in X-ray observations 
of small-scale outflows in radio galaxies (\citealt{tom10}). The jet efficiency ($\epsilon$; our fiducial value is $6\times 10^{-5}$) accounts 
for both the fraction of the in-falling mass at the inner boundary (at 1 kpc for the cluster runs) 
that is accreted by the SMBH and for the fraction of accretion energy that is channeled into the jet 
kinetic energy. Our results are insensitive to a reasonable variation in jet parameters ($v_{\rm jet}$, $r_{\rm jet}$, $\theta_{\rm jet}$, $\sigma_r$, $\sigma_\theta$),
 but depend on the jet efficiency ($\epsilon$).
 
 Like \citet{gas12}, the jet energy is injected only in the form of kinetic energy; we do not add a thermal energy source term corresponding to the jet. We note 
 that \citet{li14b} have shown that the core evolution does not depend sensitively on the manner in which the feedback energy is partitioned into kinetic or 
 thermal form. Another difference from previous approaches, which use few grid points to inject jet mass/energy, is that our jet injection region is well-resolved.

\subsection{Grid, Initial \& boundary conditions}  
\label{sec:grid}

Most AGN feedback simulations evolved for cosmological timescales (e.g., \citealt{gas12,li14}) use Cartesian grids with mesh refinement.
However, we use spherical coordinates with a logarithmically spaced grid in radius, and equal spacing in $\theta$ and $\phi$. The advantage 
of a spherical coordinate system is that it gives fine resolution at smaller scales without a complex algorithm. Perhaps more importantly, a 
spherical set up allows for 2-D axisymmetric simulations which are much faster and capture a lot (but not all) of essential physics.

We perform our simulations in spherical coordinates with $0 \leq \theta \leq \pi$, $0 \leq \phi \leq 2 \pi$, and  $r_{\rm min} \leq r \leq$ $r_{\rm max}$, with
$$
r_{\rm [min,max]} = [1,~200]~{\rm kpc} \left ( \frac{M_{200}}{7\times 10^{14} \msun} \right)^{1/3}.
$$ 
According to self similar scaling, we have scaled all length scales in our simulations (inner/outer radii $r_{\rm min}/r_{\rm max}$,  $r_{200}$, 
jet radius $r_{\rm jet}$) as $M_{200}^{1/3}$. 

We apply outflow boundary conditions ({gas is allowed to leave the computational domain but prevented from entering it}) at the inner radial boundary. 
We fix the density and pressure at the outer radial boundary to the initial value and prevent gas from leaving or entering through the outer boundary.
Reflective boundary conditions are applied in $\theta$ (with the sign of $v_\phi$ flipped) and periodic boundary conditions are used in $\phi$.
We noticed that cold gas has a tendency to artificially `stick' at the $\theta$ boundaries (mainly in 2-D axisymmetric simulations) for our 
reflective boundary conditions. This cold gas can lead to 
an unphysically large accretion rate close to the poles, and hence artificially enhanced feedback heating (Eq. \ref{eq:fb}). Therefore, we exclude 8 
grid-points at each pole 
when calculating the mass accretion rate; these excluded angles correspond to only 0.5\% of the total solid angle for 128 grid points 
in the $\theta$ direction. All our diagnostics ($\dot{M}_{\rm acc}$, entropy profiles, etc.) also exclude these small solid angles close to the poles.

The resolution for $3-D$ runs is $256\times128\times32$ and for $2-D$ runs is $512\times256$. Since we use a logarithmic grid in the radial direction, the 
resolution for 256 (512) grid points in the radial direction corresponds to a good resolution of $\Delta r/r=0.02~(0.01)$. {The minimum resolution 
in the radial direction for the fiducial 3D (2D) run is $\approx 0.02~(0.01)$ kpc.} For such a resolution our 
integrated quantities (mass accretion rate, jet power, cold gas mass, etc.) are converged.

We focus on simulations of a galaxy cluster with $M_{200}=7 \times 10^{14} \msun$ but with different parameters such as feedback efficiency. 
For comparison we also carried out simulations for a massive cluster with $M_{200}=1.8\times 10^{15} \msun$. 
The initial conditions are the same as in \citet{sha12}; i.e., we assume the initial entropy profile ($K\equiv T_{\rm keV}/n_e^{2/3}$; $T_{\rm keV}$ is the ICM 
temperature in keV and $n_e$ is the electron number density) of the form
\begin{equation}
\label{eq:ent}
K(r) = K_0 + K_{100} \left ( \frac{r}{100~{\rm kpc}} \right)^{1.4},
\end{equation}
as suggested by \citet{cav09}.\footnote{Whether an entropy core exists is debated (\citealt{pan14}), but our results are insensitive to our initial conditions. 
Our ICM profiles change with time and reach a quasi-steady state which may or may not have an entropy core.} 
For our cluster runs, we set $K_0 = 10$ keV cm$^2$ and $K_{100} = 110$ keV cm$^2$  at the start (as in \citealt{sha12}).
We assume self-similar behavior scaling with $M_{200}$
(\citealt{kai86}) to set the initial entropy profile for our massive cluster runs (i.e. we assume $K_0 =19$ keV cm$^2$ and  $K_{100} =210$ keV cm$^2$; 
c.f. \citealt{mcc08}).
Except for early transients, our results are independent of the precise choice of the initial values of $K_0$ and $K_{100}$.

The outer electron number density is 
fixed to be $n_e=0.0015$ cm$^{-3}$. Given the entropy profile and the density at the outer radius, we can solve for the hydrostatic density and 
pressure profiles in an NFW potential (Eq. \ref{eq:NFW}). We introduce small (maximum overdensity is 0.3) isobaric density perturbations on top of
 the smooth density (for details, see \citealt{sha12}). 

\begin{table*}
\caption{Table of runs}
{\centering
\resizebox{0.9 \textwidth}{!}{%
\begin{tabular}{ccccccccc}
\hline
\hline
Label & dim. & $N_r\times N_\theta \times N_\phi$ & min. resolution &$M_{200}$ & jet efficiency ($\epsilon$)  & $ \lan \dot{M}_{\rm acc} \ran^{\ddag} $ & $\lan \dot{M}_{\rm acc,hot} \ran / \lan \dot{M}_{\rm acc,cold} \ran $ & jet duty \\ 
 & & & (kpc) & ($\msun$) & & $(\msun {\rm yr}^{-1})$ & &  cycle$^{\dag\dag}$ (\%) \\
\hline
C6m5D3$^\dag$ & 3 & $256\times128\times32$ & 0.02 & $7\times10^{14}$ & $6\times 10^{-5}$ & 25.1 (244.2)$^\ddag$ & 0.2 (0.06) & 59.6 \\ 
C5m4D3 & 3 & $256\times128\times32$ & 0.02 & $7\times10^{14}$ & $5\times 10^{-4}$ & 7 & 0.78 & 82 \\
C1m2-D3 & 3 & $256\times128\times32$ & 0.02 & $7\times10^{14}$ & $0.01$ & 1.9 & 1.3 &99.8 \\
C6m5D2$^\dag$ & 2 & $512\times256\times1$ & 0.01 & $7\times10^{14}$ & $6\times 10^{-5}$ & 23.5 (170)$^{\ddag}$ & 0.23 (0.19) &64.8 \\ 
C6m6D2 & 2 & $512\times256\times1$ & 0.01 &$7\times10^{14}$ & $6\times 10^{-6}$ & 153 & 0.27 & 47.3\\
C1m2-D2 & 2 & $512\times256\times1$ & 0.01 & $7\times10^{14}$ & 0.01 & 0.77 & 14.8 & 99.8 \\
C1m4D2 & 2 & $512\times256\times1$ & 0.01 & $7\times10^{14}$ & $10^{-4}$ & 13.7 & 0.3 & 63.8\\
C5m4D2 & 2 & $512\times256\times1$ & 0.01 & $7\times10^{14}$ & $5 \times 10^{-4}$ & 5.1 & 1.6 & 72.9 \\
M6m6D2 & 2 & $512\times256\times1$ & 0.014 & $1.8\times10^{15}$ &  $6 \times 10^{-6}$ & 293 (299) & 0.3 (0.5) & 0.0\\
M6m5D2 & 2 & $512\times256\times1$ & 0.014 & $1.8\times10^{15}$ &  $6 \times 10^{-5}$ & 77.7  & 0.58 & 50.1\\
M1m4D2 & 2 & $512\times256\times1$ & 0.014 & $1.8\times10^{15}$ &  $10^{-4}$ & 48.18  & 0.62  & 47.1\\
M5m4D2 & 2 & $512\times256\times1$ & 0.014 & $1.8\times10^{15}$ &  $5 \times 10^{-4}$ & 18.7  & 2.9  & 63.8 \\
M1m2-D2 & 2 & $512\times256\times1$ & 0.014 & $1.8\times10^{15}$ &  $0.01$ & 8.1 & $3.7\times10^5$  & 99.8\\
\hline
\end{tabular} }\\}
{\textbf{Notes}} \\
`C' in the label stands for a cluster ($M_{200}=7 \times 10^{14} \msun$) and `M' for a massive 
cluster ($M_{200}= 1.8 \times 10^{15} \msun$). Label C6m5D3 indicates that it is a cluster run in 3-D with an efficiency $\epsilon=6\times 10^{-5}$ (Eq. \ref{eq:fb}).\\
{$^\dag$}{The fiducial 3-D and 2-D  runs.} \\
{$^\ddag$}{Angular brackets denote time average over the full run.  The quantities in brackets denote values for a pure 
cooling flow ($\dot{M}_{\rm cf}$). Note that 8 grid points close to the poles are excluded when calculating the accretion rates.}\\
{$^{\dag\dag}$}{Jet duty cycle is defined as the fraction of total time for which the jet power is $>10^{40}$ erg s$^{-1}$.}\\
\label{table}
\end{table*}

\section{Results}
\label{sec:res}
In this section we describe the key results from our simulations. Table \ref{table} lists our runs. 
We begin with the results from our fiducial 3-D cluster run (C6m5D3 in Table \ref{table}). We show that the 1-D profiles of density, entropy, etc.
 are consistent with observations. We highlight the cycles of cooling and AGN jet 
feedback, and the spatial and velocity distribution of the cold gas. We show that there are three components in cold ($T<5\times 10^4$ K) 
gas distribution: a massive, 
centrally-concentrated, rotationally-supported torus; spatially extended and fast ($\gtrsim 500$ km s$^{-1}$) outflows correlated with jets; and slower 
($\lesssim 300$ km s$^{-1}$) in-falling cold gas that condenses out because of local thermal instability. Then we compare the results from our 3-D and 
2-D axisymmetric simulations. We also explore the dependence of our results on the halo mass and the jet efficiency.

\begin{figure*}
\plotone{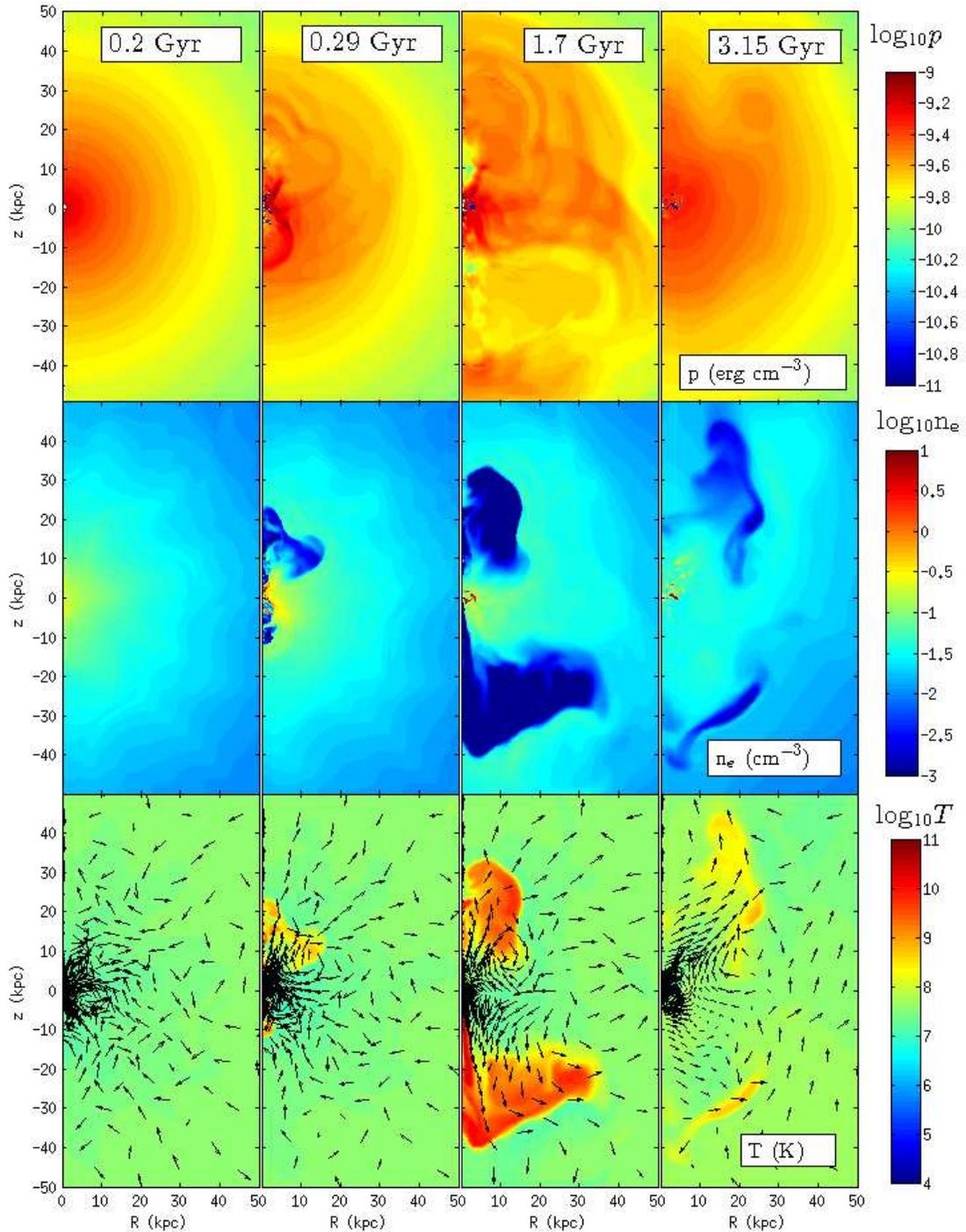}
  \caption{The {pressure (upper panel), electron number density (middle panel), and temperature (lower panel)} contour plots ($R-z$ plane at $\phi=0$) in the core at different times for the 3-D fiducial run. The density is cut-off at the maximum and the minimum contour level shown. The low-density bubbles/cavities are not symmetric and there are signatures of mixing in the core. The left panel corresponds to a time just before a cooling time in the core. The second panel from left shows cold gas dredged up by the outgoing jets. The rightmost panel shows in-falling extended cold clouds. 
{The pressure maps show the weak outer shock, but the bubbles/cavities so prominent in the density/temperature plot are indiscernible in the pressure map, 
implying that the bubbles are in pressure equilibrium and buoyant. Also notice the outward-propagating sound waves in the two middle pressure panels in which the 
jet is 
active. The in-falling/rotationally-supported cold gas has a much lower temperature and pressure than the hot phase. The arrows in the temperature plots 
denote the projected gas velocity unit vectors.}}
  \label{fig:3-D_fiducial}
\end{figure*}
               
               
\subsection{The fiducial 3-D run}

We experimented with different values of jet efficiencies ($\epsilon$; Eq. \ref{eq:fb}) in our 3-D cluster ($M_{200}=7 \times 10^{14} \msun$) simulations, and found 
that the average mass accretion rate for 
$\epsilon=6 \times 10^{-5}$ was about 10\% of a pure cooling flow (see Table \ref{table}). Therefore, we choose this as our fiducial value, which is 
smaller compared to 
the values chosen by some recent works (\citealt{gas12,li14,li14b}), but is consistent with observational constraints (e.g., \citealt{ode08}). Our fiducial value 
should be considered as the smallest efficiency that is required to prevent a cooling flow in a cluster (this critical efficiency depends on the halo mass, as 
we shall see later).

The minimum ratio of the cooling time ($t_{\rm cool} \equiv 3nk_B T/[2 n_e n_i \Lambda]$) and the local free-fall time 
($t_{\rm ff} \equiv [2r/g]^{1/2}=[2 r^3/GM(<r)]^{1/2}$) 
is 7 for the initial ICM; this ratio (\tr) is a good diagnostic of the state of the cluster core in rough thermal balance. Since the initial condition is in 
hydrostatic equilibrium, there is negligible accretion 
through the inner boundary, and therefore there is no jet injection. However, after a cooling time in the core ($\approx 200$ Myr) there is a rise in the accretion 
rate across the inner boundary ($\dot{M}_{\rm acc}$), and hence in jet momentum injection (Eq. \ref{eq:fb}). 
The jet powers a bubble that heats the core and raises \tr, keeping the mass
accretion rate well below the cooling flow value (c.f. top panel of Fig. \ref{fig:2-D3-D_vs_time}). After this time the cluster core is in a state of average global thermal 
balance between radiative cooling and feedback heating via AGN jets.

\subsubsection{Jets, bubbles \& multiphase gas}

Figures \ref{fig:3-D_fiducial} show the snapshots ($r-\theta$ plane at $\phi=0$) of pressure, density, {and temperature} at different times for our fiducial 3-D run. The X-ray emitting ICM 
plasma is quite distinct from the dense cold ($10^4$ K) gas and from the low-density jet/bubble. The cold gas accreting on to SMBH gives rise to AGN jets.
Before a cooling time (0.2 Gyr) there are no  signs of cooling and jets. After a cooling time, accretion rate through the inner boundary (at 1 kpc) increases 
and bipolar jets are launched (0.29 Gyr). The jets are not perfectly symmetric, as they are shaped by the presence of cold gas in their way. The 
inhomogeneities in the ICM enhances mixing with (and stirring of) the ICM core, resulting in effectiveness of our jets even with a low efficiency. 

Jets are fast in the injection region but become slow, buoyant, and almost in pressure balance with the ICM (compare the upper and middle panels of 
Fig. \ref{fig:3-D_fiducial} ) because of turbulent drag and sweeping up of 
the ICM. In absence of further power injection, the bubbles are detached from the jets and rise buoyantly and mix with the ICM at 10s 
of kpc scales (3.15 Gyr in Fig. \ref{fig:3-D_fiducial}). Most of the cold gas is very 
centrally concentrated (within 10 kpc), but does condense out at larger radii, although never beyond 30 kpc. 

 As jets plough through the dense cold gas clouds, forward shock moves ahead of these clouds after partially disrupting them. The collision 
 results in a reverse shock and a huge back-flow of hot jet material which mixes with the cooler ICM, driving the core entropy to higher values. These 
 back-flows and mixing are mainly responsible for heating the cluster core.

\subsubsection{Radial profiles} 

\begin{figure*}
\plotone{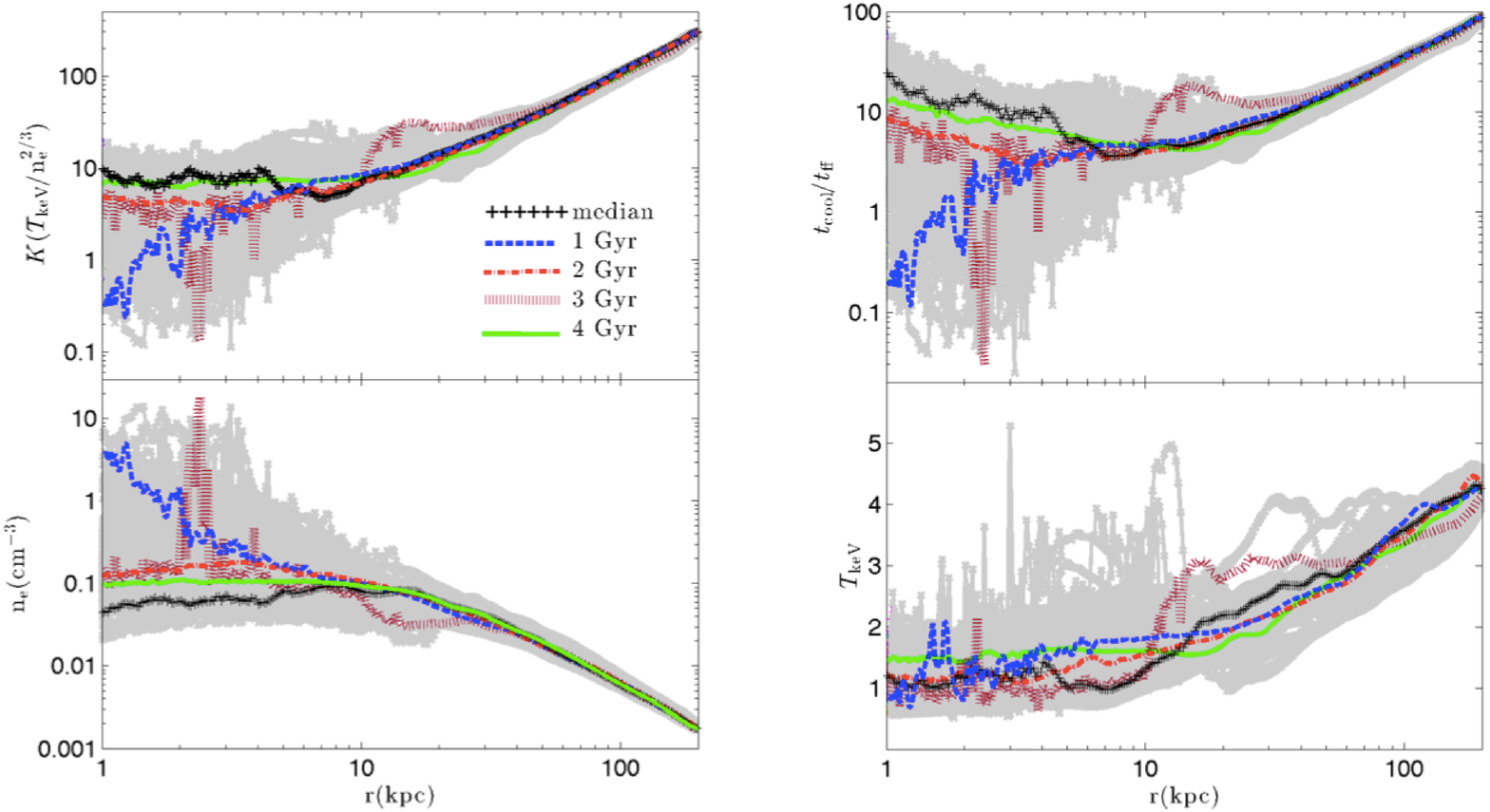}
  \caption{X-ray emissivity-weighted (considering only 0.5-8 keV gas) 1-D profiles of important thermodynamic variables as a function of radius. Snapshots
  at 1, 2, 3, 4 Gyr are shown. Various
  quantities are obtained by combining 1-D profiles of density and pressure. The median and standard deviation ($\sigma$) of entropy 
  ($K\equiv T_{\rm keV}/n_e^{2/3}$) at 
  20 kpc are calculated.  Various profiles corresponding to the median entropy at 20 kpc  (14 keV cm$^2$) are shown in different panels 
  (black lines with `+'). Thick grey lines show the profiles for which the entropy at 20 kpc is within $1-\sigma$ of its median value. }
  \label{fig:1D_profiles}
\end{figure*}

Before discussing the detailed kinematics of cold gas and jet cycles, we show in Figure \ref{fig:1D_profiles} the 1-D profiles of important thermodynamic 
quantities (entropy $[T_{\rm keV}/n_e^{2/3}]$, $t_{\rm cool}/t_{\rm ff}$, $n_e$, $T_{\rm keV}$) as a function of radius for the fiducial 3-D run. In addition to
the instantaneous profiles (at 1, 2, 3, 4 Gyr), the median profile and spread about it are shown. The median is calculated 
for the entropy measured at 20 kpc ({roughly the core size}) and all the profiles with entropy within one standard deviation at the same radius are shown in grey. 

The spread in quantities outside $\sim 20$ kpc is quite small, but increases toward the center {because multiphase cooling (leading to density spikes) 
and strong jet feedback (leading to overheating) are most effective within the core.} 
{The density at 1 Gyr is peaked toward the center, indicating that the cluster core is in a cooling phase.}
The spikes in density at 
3 Gyr have corresponding spikes in entropy and $t_{\rm cool}/t_{\rm ff}$ profiles, but not as prominent in the temperature profile. The temperature 
fluctuations are rather modest compared to fluctuations in other quantities because of dropout and adiabatic cooling. Temperature profiles show a 
general increase with radius, as seen in observations.
 
There is a large spread in entropy toward lower values about the median at radii $< 10$ kpc (top-left panel in Figure \ref{fig:1D_profiles}). This is because
there are short-lived cooling events during which the entropy in the core decreases significantly (simultaneously, density increases and 
$t_{\rm cool}/t_{\rm ff}$ decreases). On the other hand, the increase in the core entropy is smaller but lasts for a 
cooling time, which is longer in this state. This behavior is generic, fairly insensitive to parameters such as the feedback efficiency and the halo mass.

\subsubsection{The cold torus} 

\begin{figure*}
 \plotone{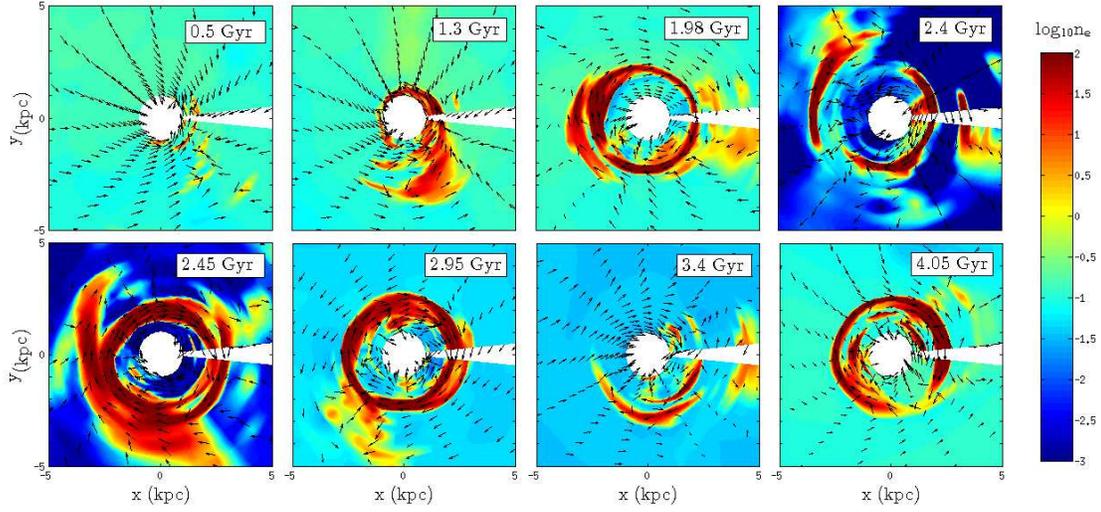}
  \caption{The 2-D ($z=0$) contour plots of electron number density (in cm$^{-3}$) in the mid-plane of the very inner region at different times for the fiducial 3-D run, 
  with the projection of the velocity 
  unit-vector represented by arrows. 
  The top-left panel shows the beginning of the infall of cold gas with random angular momentum. The {top-second left} panel shows an anti-clockwise transient torus.
  All times after this show a clockwise torus in the mid-plane which waxes and wanes because of cooling and AGN heating cycles. 
  Even at late times the cold torus is not stable 
  and gets disrupted by jets.}
  \label{fig:3-D_torus}
\end{figure*}

While Figure \ref{fig:3-D_fiducial} shows that cold gas can be dredged up by AGN jets (second panel; see also \citealt{rev08,pop10}) and can also condense out of the ICM at large scales 
(fourth panel), majority of cold gas is at very small scales ($< 5$ kpc) in the form of an angular-momentum supported cold torus. Figure \ref{fig:3-D_torus} shows
the zoomed-in density snapshots in the equatorial ($\theta=\pi/2$) plane at different times; the arrows show the projection of velocity unit vectors.
As the cluster evolves the cold gas, condensing out of the hot ICM, gains angular momentum from jet-driven turbulence. Because of a significant angular 
velocity, an angular momentum barrier forms and cold gas circularizes at small radii. 

Unlike \citet{li14b}, our cold torus is dynamic in nature as AGN jets disrupt it time and again, but it reforms due to cooling. Figure \ref{fig:3-D_torus}
 shows the evolution of the torus at various stages of the simulation. The top-left panel of Figure \ref{fig:3-D_torus} shows the cluster center at 0.5 Gyr. Small 
 cold gas clouds are accumulating in the core after the first active AGN phase. 
 At 1.3 Gyr, cold gas accreting through the inner boundary has an anti-clockwise rotational sense.  
 At 1.98 Gyr, cold gas (and the hot gas out of which it condenses) is rotating clockwise. Jet activity leading up to this phase has 
 reversed the azimuthal velocity of the cold gas.  
 At all times after this the dynamic cold gas torus rotates in a clockwise sense, essentially because the mass (and angular momentum) in the rotating torus is 
 much larger than the newly condensing cold gas. 

The torus gets disrupted due to jet activity but forms again quickly. The snapshots at 2.4 and 2.45 Gyr show that the inner region is covered by
the very hot/dilute jet material. If the jets were rapidly changing direction as argued by \citet{bab13}, we would in fact expect the cold gas torus to be occasionally 
disrupted by the jets.  In the present simulations, however, this behavior is an artifact of our feedback prescription;
we scale the jet power with the instantaneous 
mass inflow rate through the inner boundary (see Eq. \ref{eq:fb}). Even small oscillations of the cold torus can sometimes lead to a large instantaneous 
mass inflow through the inner boundary and hence an explosive jet event. The reassuring fact is that
these explosive `events' are rare and the jet material is quickly mixed with the ICM after these.
In reality, most of the cold gas in 
the torus will be consumed by star-formation. Only the low angular momentum cold gas that circularizes closer in ($\lesssim 100$ pc) can be accreted
by the SMBH at a short enough timescale.

{A cold torus forms in all our 3D cluster simulations with different efficiencies. However extended 
cold gas is lacking at late times in simulations with high jet efficiencies.}
\citet{li14b} show that after 3 Gyr the cold gas settles down in form of a stable torus, with no further condensation of extended cold gas. This is inconsistent 
with observations which show that about a third of cool-core clusters show H$\alpha$ filaments extending out to 10s of kpc from the center (\citealt{mcd10}).
The bottom panels in Figure \ref{fig:3-D_torus} from our fiducial run show that the torus is unsteady even at late times with 
extended cold gas condensing out till the end of our run. We compare our results in detail with \citet{li14b} in section \ref{sec:comp}. 

\subsubsection{Velocity and space distribution of cold gas}

\begin{figure*}
\plotone{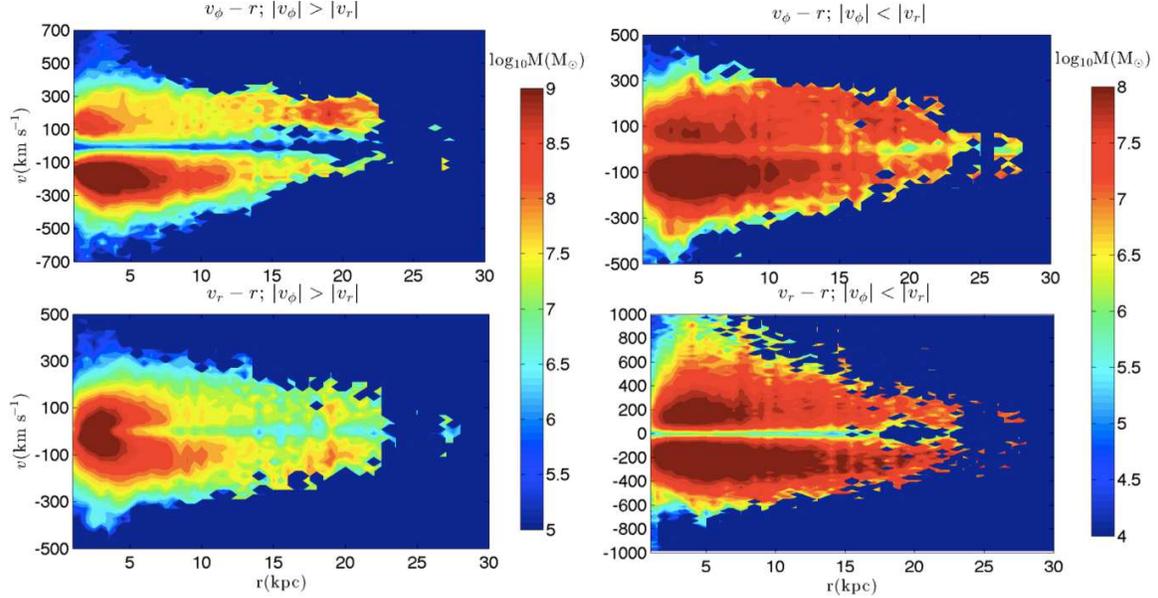}
  \caption{The velocity-radius distribution of the cold gas ($T<5 \times 10^4$ K) mass averaged from 1 to 4 Gyr. The top-left panel shows the $v_\phi-r$ mass 
  distribution ($\frac{d^2M}{d\ln |v_\phi| d \ln r}$; $\Delta v_\phi=\Delta v_r=20~{\rm km~s}^{-1},~\Delta r=0.5~{\rm kpc}$ are the bin-sizes) for 
  the rotationally-dominant ($|v_\phi|>|v_r|$) gas and 
  the bottom-left panel shows the $v_r-r$ distribution for the same gas. The top-right panel shows the $v_\phi-r$ distribution 
  for the radially-dominant 
  ($|v_r|>|v_\phi|$) cold gas and the bottom-right panel shows the $v_r-r$  ($\frac{d^2M}{d\ln |v_r| d \ln r}$) 
  distribution for the same gas. Some of the salient features are:
  the rotationally-dominant cold gas, which is concentrated mainly within 5 kpc, is more abundant by a factor $\sim 40$ than the radially-dominant gas; 
  the dominant rotationally-supported clockwise cold torus with a negligible radial velocity (see Fig. \ref{fig:3-D_torus}) is clearly visible in the two left panels; 
  the radially-dominant
  cold gas (with $|v_r|>|v_\phi|$) is much more radially extended, going out to 25 kpc; the bottom-right panel shows that the in-falling ($v_r<0$) cold gas dominates over the outgoing 
  cold gas (by a factor $\approx 2$) and that the outgoing cold gas at $\lesssim 5$ kpc extends to very large velocities.}
  \label{fig:vel_r_phase}
\end{figure*}

We find it very instructive
to classify the cold gas into two components: most of the mass is in the {\em rotationally-dominant} gas at $\lesssim 5$ kpc; a smaller fraction is in a {\em radially 
dominant} component spread over 20 kpc. Figure \ref{fig:vel_r_phase} shows the velocity and space distribution of rotationally (the left two panels; 
$d^2M/d\ln |v_\phi| dr$ and $d^2 M/d \ln |v_r| dr$; $|v_\phi|>|v_r|$) and radially (the right two panels; 
$d^2M/d\ln |v_\phi| dr$ and $d^2 M/d\ln |v_r| dr$; $|v_\phi|<|v_r|$) dominant cold 
($T<5 \times 10^4$ K) gas, averaged from 1 to 4 Gyr. The rotationally dominant gas distribution ($|v_\phi| > |v_r|$; two left panels in Fig. \ref{fig:vel_r_phase}) 
shows two peaks at $v_\phi \approx \pm 150-200$ km s$^{-1}$ and $r \approx 1.5-4$ kpc, corresponding to the cold tori seen in Figure \ref{fig:3-D_torus}. The 
radial velocity is $\lesssim 100$ km s$^{-1}$.

The distribution of the radially dominant cold gas in Figure \ref{fig:vel_r_phase} is quite different from the rotationally dominant gas. In addition to 
a larger radial extent, the radial velocity of
the radially dominant component is much larger, going up to $\pm 600$ km s$^{-1}$, much larger than the maximum azimuthal speed. The radial velocity 
of the closer in gas ($\lesssim 5$ kpc) is 
even larger for the outflowing ($v_r>0$) component because it is dredged up by the fast jet material; tiny mass in the cold gas is seen to reach a velocity close 
to  $v_{\rm jet}=0.1 c$. The mass in the in-falling radially-dominant cold gas is $\approx$ twice that of the outgoing cold gas. 

\begin{figure}
\epsscale{1.2}
\plotone{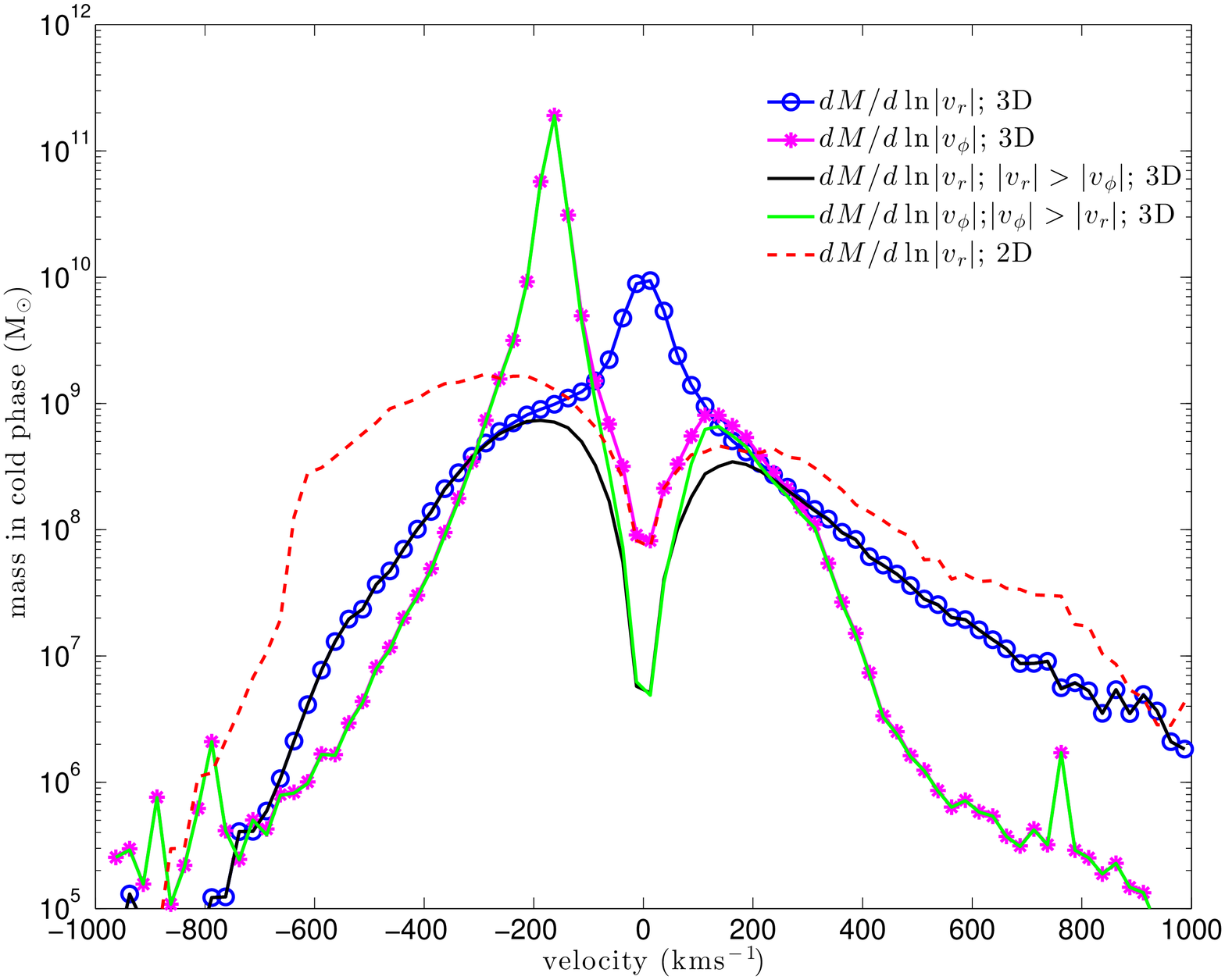}
\caption{The velocity distribution of cold gas for the 3-D fiducial run with respect to the radial and azimuthal velocities. Also shown are the rotationally 
($|v_\phi|>|v_r|$) and radially ($|v_r|>|v_\phi|$) dominant components. At large velocities the total and radially/rotationally dominant components coincide 
but at small velocities they do not, as expected (at low velocities the other component of velocity dominates the mass budget). 
Also shown (dashed line) is the radial velocity distribution for the 2-D fiducial run; the azimuthal velocity 
is zero for 2-D axisymmetric runs. 
}
\label{fig:velpdf}
\end{figure}

Figure \ref{fig:velpdf} shows the 1-D velocity distribution of the cold gas averaged from 1 to 4 Gyr. The two large, sharp peaks correspond to the 
massive clockwise rotating cold torus. The radially-dominant component ($|v_r|>|v_\phi|$) shows a prominent high velocity tail in the positive direction.
The negative velocity component for velocities larger than 300 km s$^{-1}$ is also dominated by the radially in-falling (rather than rotationally dominant) 
gas, sometimes affected by the fast jet back-flows. The maximum velocity peak of the radially and rotationally dominant cold gas coincide at $\approx 150-200$ 
km s$^{-1}$, corresponding to the circular velocity at $\sim 5$ kpc.

\begin{figure}
\psfrag{e}[r][r][1.2][0]{time(Gyr)}
\psfrag{f}[c][c][1.2][0]{normalized quantities}
\psfrag{a}[r][r][0.7][0]{$\dot{M}_{\rm in, cold, 5 kpc} (M_\sun {\rm yr}^{-1})$}
\psfrag{b}[r][r][0.7][0]{$\dot{M}_{\rm out, cold, 5 kpc} (M_\sun {\rm yr}^{-1})$}
\psfrag{c}[r][r][0.7][0]{$\dot{M}_{\rm acc} (M_\sun {\rm yr}^{-1})$}
\psfrag{ddddddddddddddddddddddddd}[r][r][0.7][0]{jet power ($10^{41}$ erg s$^{-1}$)}
\epsscale{1.2}
\plotone{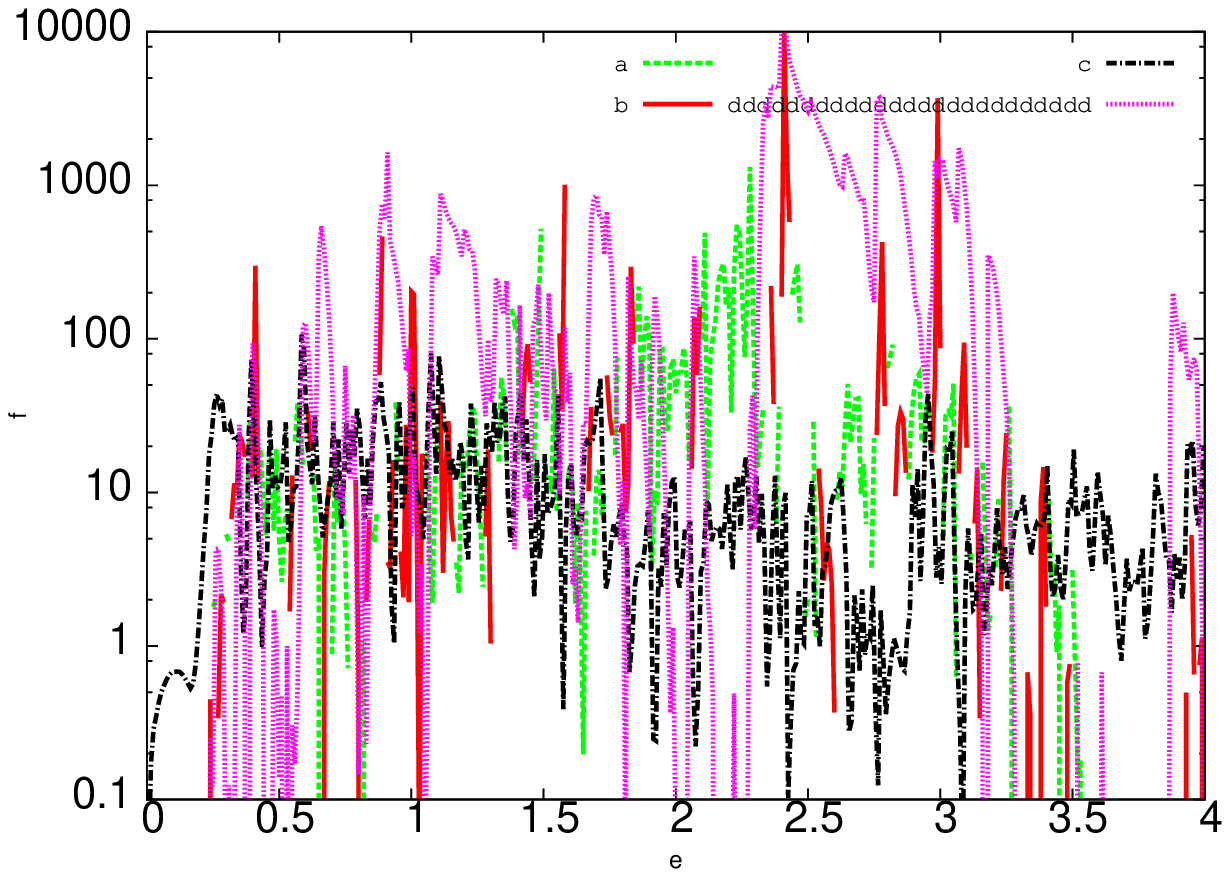}
\caption{The mass inflow (green short-dashed line) and outflow (red solid line) rates in the cold phase measured at 5 kpc as a function of time. Also shown are 
the jet power (normalized to $10^{41}$ erg s$^{-1}$; how jet power is calculated is described in section \ref{sec:cycles}) and the mass accretion rate at 
1 kpc (Eq. \ref{eq:fb}). Note that the largest spikes in the cold outflow rates are 
mostly associated with a sudden rise in jet energy, indicating cold gas uplifted by jets. The difference between the mass inflow rate at 5 kpc and at 1 kpc 
(clearly noticeable at $\sim 2$ Gyr) leads to the build-up of the rotating cold torus (see Fig. \ref{fig:3-D_torus}). Also note that extended cold gas 
and jets are absent at $\sim 3.75$ Gyr.}
\label{fig:jet_in_out}
\end{figure}

Figure \ref{fig:jet_in_out} shows the relationship of in-falling and outgoing cold gas at small scales (5 kpc) and AGN jet activity. The cold outflow rate 
shows large spikes coincident with a sudden rise in AGN jet power, implying that cold gas observed with large velocity (inf Figs. \ref{fig:vel_r_phase} 
\& \ref{fig:velpdf}) is dredged up by fast moving jets. The coincidence
is particularly strong when a massive cold gas torus is present at small scales. For steady cooling in absence of angular momentum, we expect the mass inflow
rate at 5 kpc and 1 kpc to closely follow each other. This is, however, not the case (especially around 2 Gyr) when majority of in-falling cold gas crossing 5 kpc
is incorporated in the rotating cold torus, instead of accreting through 1 kpc. Also note that the outflowing cold gas is in form of very short-lived massive spikes, 
but the inflowing cold gas is smoother. The interpretation is that the outflowing cold gas is associated with the AGN-uplifted cold torus gas, 
and the infalling cold gas is because of local thermal instability in a gravitational field.

\subsubsection{Cooling \& heating cycles}
\label{sec:cycles}

\begin{figure*}
\epsscale{1.0}
\plotone{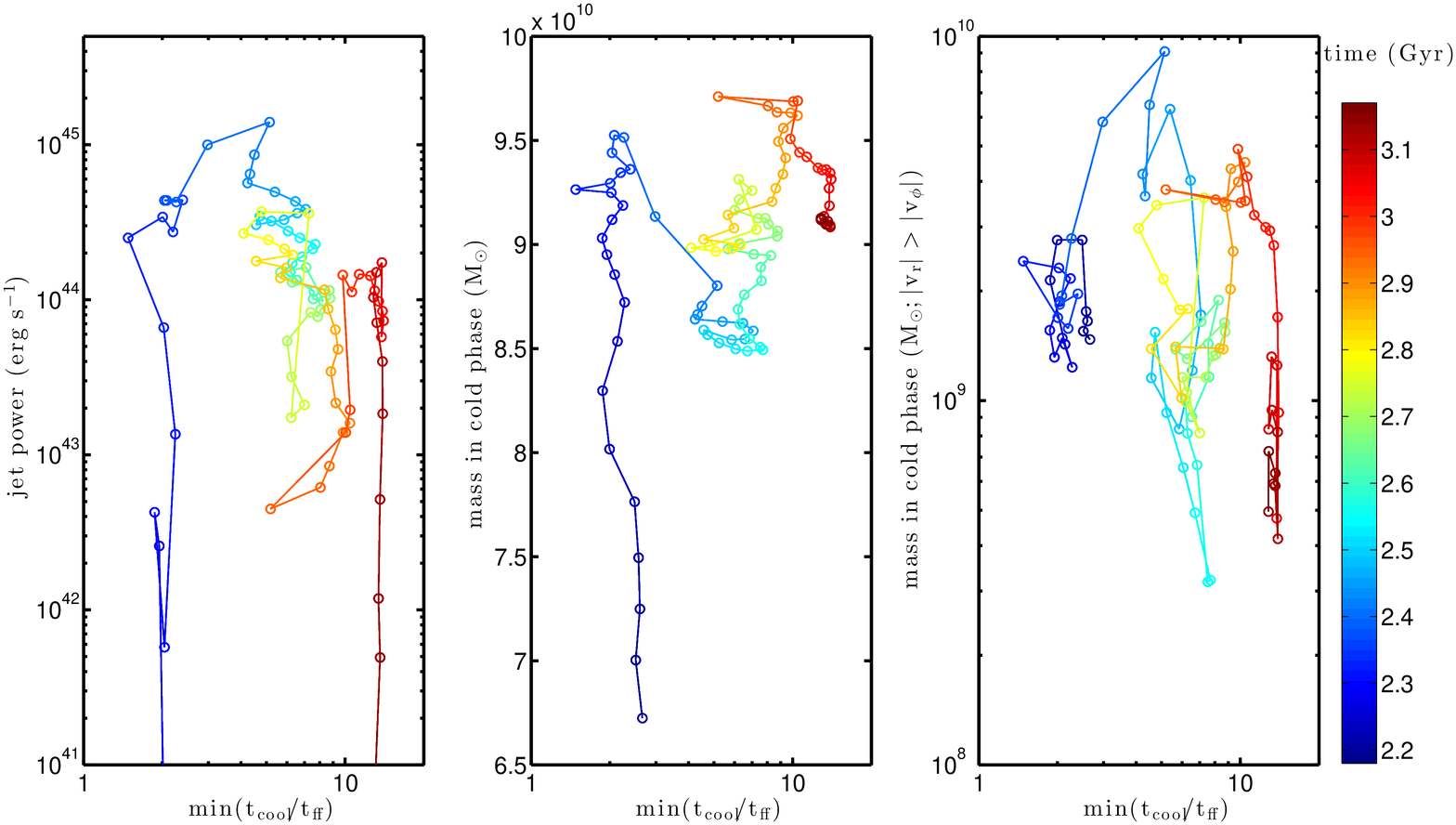}
\caption{The variation of jet power, total cold gas mass and the radially-dominant cold gas mass as a function of min(\tr) for the fiducial 3-D run from 2.2 to 3.2 Gyr. Color shows 
the evolution of cluster in time. While jet power ({left panel}) and radially-dominant cold gas mass ({right panel}) show clockwise cycles with min(\tr), the total cold gas mass ({middle panel}) simply builds up in time. {Notice the linear scale for the total cold gas mass instead of a log scale for
 the other two cases.}
}
\label{fig:3-D_cycle}
\end{figure*}

One of the distinct features of the cold feedback paradigm is that we expect correlations in jet power, cold gas mass, mass accretion rate, min(\tr), core 
entropy, etc. The observations indeed show such correlations (e.g., Figs. 1, 2 in \citealt{cav08}; see also \citealt{voi14,sun09,mcd11}). In Figure 
\ref{fig:3-D_cycle} we make 
`phase-space' plots of jet power and cold-gas mass (total and the radially-dominant component) as a function of min(\tr) for our fiducial 3-D run. 

{\em Evaluating min(\tr):} The ratio \tr~is calculated by making radial profiles of emissivity-weighted (only including plasma in the range of 0.5 to 8 keV) internal 
energy and mass densities. They are 
combined to calculate $t_{\rm cool} \equiv (3/2) n k_B T/(n_en_i \Lambda)$, and \tr~is calculated by taking its ratio with the free-fall time based on the NFW 
potential (Eq. \ref{eq:NFW}; $t_{\rm ff} \equiv [2r/g]^{1/2}$ where $g \equiv d\Phi/dr$). The broad local minimum in \tr~profile is searched going in 
from the outer radius and is used as min(\tr). 


{\em Evaluating jet power:} The jet power is also calculated in a novel way, which is close to what 
is done in observations.\footnote{Observers calculate the bubble/cavity mechanical power by assuming it to be in pressure 
balance with the background ICM and by using a size and an age estimate for the bubble 
(e.g., see \citealt{bir04}). Indeed, our bubbles are in pressure balance with the ICM, as seen in Fig. \ref{fig:3-D_fiducial}.} 
We consider the grids with mass density lower than a threshold value (chosen to be 0.17 times the initial minimum density
 in the computational volume; results are insensitive to the exact value of the threshold density) to belong to the jet/bubble material, and we simply 
 volume-integrate the internal energy density of all such cells to calculate the jet energy (only considering thermal energy; 
 we use $\gamma=5/3$ for the jet material because it is a
 non-relativistic hot gas in our simulations; in reality, relativistic particles with $\gamma=4/3$ are a 
major component of jets). This density-based definition of the jet material coincides with the visual appearance of the jet. The jet energy is divided by
an estimate of the bubble lifetime (chosen to be 30 Myr, of order the dynamical/buoyancy time at 10 kpc; see Table 3 in \citealt{bir04})  to arrive 
at the jet power. For simplicity, we use the same value of the bubble lifetime at all times in all our runs. A trivial definition of jet power, in which it is
proportional to the instantaneous accretion rate at 1 kpc, is given by Eq. \ref{eq:fb} as $3.4 \times 10^{42} \dot{M}_{\rm acc}(\msun {\rm yr}^{-1})$ erg s$^{-1}$.
Assuming this conversion, Figure \ref{fig:jet_in_out} shows that the two estimates of jet power are comparable 
in magnitude but vary rather differently with time. This is because, while $\dot{M}_{\rm acc}$ is an instantaneous quantity varying on a dynamical timescale, 
our jet power is based on the jet thermal energy which is an integrated quantity.

We anticipate cycles in the evolution of min(\tr) and jet power or the radially-dominant cold gas. 
Imagine that there is no accreting cold gas at the center; in this state without heating the core is expected to cool below \tr$\sim 10$ (because  
accretion rate in the hot mode is small). The \tr$\lesssim 10$ state is prone to cold-gas condensation and enhanced feedback heating 
if $\epsilon$ is sufficiently high. Energy injection leads to overheating of 
the core and an increase in \tr; since condensation/accretion is suppressed in this state, both jet power (because of adiabatic/drag losses) 
and radially-dominant cold gas mass are reduced in this state of \tr$>10$. Eventually the core cools again and the cycle starts afresh.

The left panel of Figure \ref{fig:3-D_cycle} shows {one of the many} jet cycles in our fiducial 3-D cluster run. On average jet power vs. min(\tr)~ evolves in form of clockwise cycles of 
various  widths (a measure of the range of min[\tr]~before and after the jet event) and heights (jet power). Generally, a smaller \tr~leads to a larger mass accretion 
rate and a larger jet energy, and therefore larger overheating and a larger min(\tr). Since the efficiency of our fiducial run is rather small 
($\epsilon=6\times 10^{-5}$), the cluster core remains with \tr$< 20$ at most times. In section \ref{sec:eff} we discuss the dependence of our 
results on jet efficiency ($\epsilon$).

The middle panel of Figure \ref{fig:3-D_cycle} shows the total mass in cold gas (most of which is in the cold rotating torus) as a function of min(\tr). We see the 
mass in the cold torus building up in time. We can easily see that the {\em total} cold gas mass simply builds up in time ({see the green dashed line in the 
upper panel of Fig. \ref{fig:2-D3-D_vs_time}}), and is uncorrelated with min(\tr). 
The right panel of Figure \ref{fig:3-D_cycle} shows the mass in the radially-dominant cold gas (with $|v_r|> |v_\phi|$) as a function of
min(\tr). This panel also shows clockwise cycle like jet power shown in the left panel. A larger radially-dominant cold gas mass generally implies a 
higher accretion rate and a larger jet power, but the features in jet and cold gas cycle are not always varying in an identical fashion. {While the global evolution in phase space is clockwise, there is haphazard evolution at smaller timescales (e.g., between 2.5 to 2.9 Gyr).}


\subsection{The 2-D  runs}

The 3-D simulations are very expensive compared to the 2-D ones, not only because the number of grid cells is larger but also because the CFL time step 
is much smaller. The CFL time step in 3-D is dominated by cells close to the polar regions ($\theta=0,~\pi$) and $ \propto r \sin \theta \Delta \phi \approx 
r \Delta \theta \Delta \phi/2$, much smaller than in 2-D ($\propto r \Delta \theta$). Our $256 \times 128 \times 32$ (3-D) runs have 8 times more grid cells 
compared to our $512 \times 256$ (2-D) runs and the CFL time step is $\approx 0.2$ times smaller, making our 3-D runs 40 times more expensive than 
the 2-D ones. Therefore, for scans in various parameters (halo mass, jet efficiency, etc.), only 2-D axisymmetric simulations are practical. 
However, the key drawback is that the initially non-rotating gas cannot gain angular momentum in axisymmetry, and thus 2-D simulations do not 
show the formation of a rotationally supported torus. But, as we discuss shortly, the suppression of cooling flow, nature of radially-dominant cold gas, etc. 
are very similar in 2-D and 3-D. 

In this section we describe different variations on the fiducial setup for our 2-D simulations. Section \ref{sec:comp3-D} compares the results from 2-D and 3-D 
simulations with cooling and AGN feedback. Section \ref{sec:eff} studies the effect jet feedback efficiency and the halo mass on the properties of jets and cold gas. 
 
 \subsubsection{Comparison with 3-D}
 \label{sec:comp3-D}

Since 3-D simulations are substantially more time consuming compared to the 2-D axisymmetric ones, it will be very useful if some robust inferences can
be drawn from these faster 2-D computations. To compare the 3-D simulations with their 2-D counterparts, we have carried out the fiducial 3-D simulation in 
2-D with identical parameters (the initial density perturbations in 2-D runs are the same as the perturbation for the $\phi=0$ plane in 3-D). 

Figure \ref{fig:velpdf} compares the time-averaged velocity distribution of cold gas in 2-D and 3-D simulations. Since the azimuthal velocity vanishes in 2-D 
axisymmetric simulations, we compare the radial velocity distribution of cold gas  in 2-D simulations with the radially-dominant ($|v_r|>|v_\phi|$) component in 3-D. 
While the outflowing cold gas has a similar distribution in 2-D and 3-D, the inflowing gas is more dominant in 2-D relative to 3-D because in 3-D a lot of 
this in-falling cold gas slows down and becomes a part of the rotating cold torus.

\begin{figure}
\epsscale{1.25}
\plotone{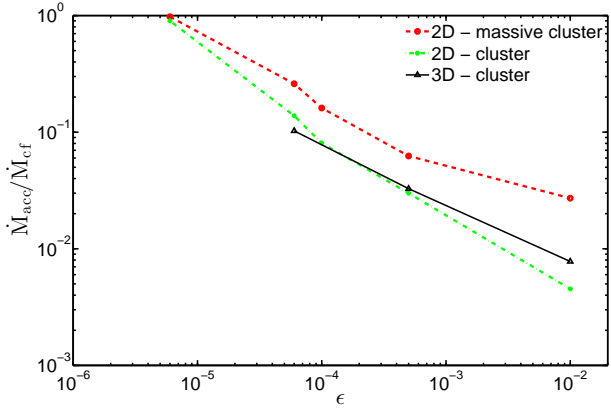}
\caption{The mass accretion rate relative to the cooling flow value as a function of jet efficiency (Eq. \ref{eq:fb}) for cluster and massive cluster runs 
(both 2-D and 3-D). The accretion rate is suppressed more for a lower halo mass at a fixed $\epsilon$.
}
\label{fig:mdotbycf}
\end{figure}

Table \ref{table} shows that the mass accretion rate through the inner boundary for the fiducial runs in 2-D and 3-D are comparable. Unlike in 3-D, we note 
 that there is substantial cold gas sticking to the poles in 2-D due to numerical reasons. Similarly, in 3-D there is a physical accumulation of cold gas 
in form of a rotating torus.

Figure \ref{fig:mdotbycf} shows the average mass accretion rate through the inner radius of our simulation volume ($\dot{M}_{\rm acc}$)
relative to the cooling flow rate ($\dot{M}_{\rm cf}$). The suppression factor ($\dot{M}_{\rm acc}/\dot{M}_{\rm cf}$) for 3-D cluster simulations (with 
$\epsilon=6\times 10^{-5},~5\times 10^{-4},~0.01$) is comparable to 2-D.

 \begin{figure}
\psfrag{a}[r][r][0.7][0]{$\dot{M}_{\rm acc}(\msun {\rm yr}^{-1})$}
\psfrag{b}[r][r][0.7][0]{cold mass ($10^9 \msun$)}
\psfrag{c}[r][r][0.7][0]{cold mass ($|v_r|>|v_\phi|;~10^8 \msun$)}
\psfrag{ddddddddddddddddddddddddddddd}[r][r][0.7][0]{${\rm min}(t_{\rm cool}/t_{\rm ff})$}
\psfrag{eeeeeeeeeeeeeeeeeeeeeeeeeeeee}[r][r][0.7][0]{jet power ($10^{42}$ erg s$^{-1}$)}
\psfrag{f}[c][c][1.2][0]{time (Gyr)}
\psfrag{g}[c][c][1.2][0]{normalized quantities, 3-D}
\psfrag{l}[r][r][0.7][0]{$\dot{M}_{\rm acc}(\msun {\rm yr}^{-1})$}
\psfrag{m}[r][r][0.7][0]{cold mass ($10^9 \msun$)}
\psfrag{nnnnnnnnnnnnnnnnnnnnnnnnnnnnnn}[r][r][0.7][0]{${\rm min}(t_{\rm cool}/t_{\rm ff})$}
\psfrag{o}[r][r][0.7][0]{jet power ($10^{42}$ erg s$^{-1}$)}
\psfrag{p}[r][r][1.2][0]{time (Gyr)}
\psfrag{q}[c][c][1.2][0]{normalized quantities, 2-D}
\epsscale{2.4}
\plottwo{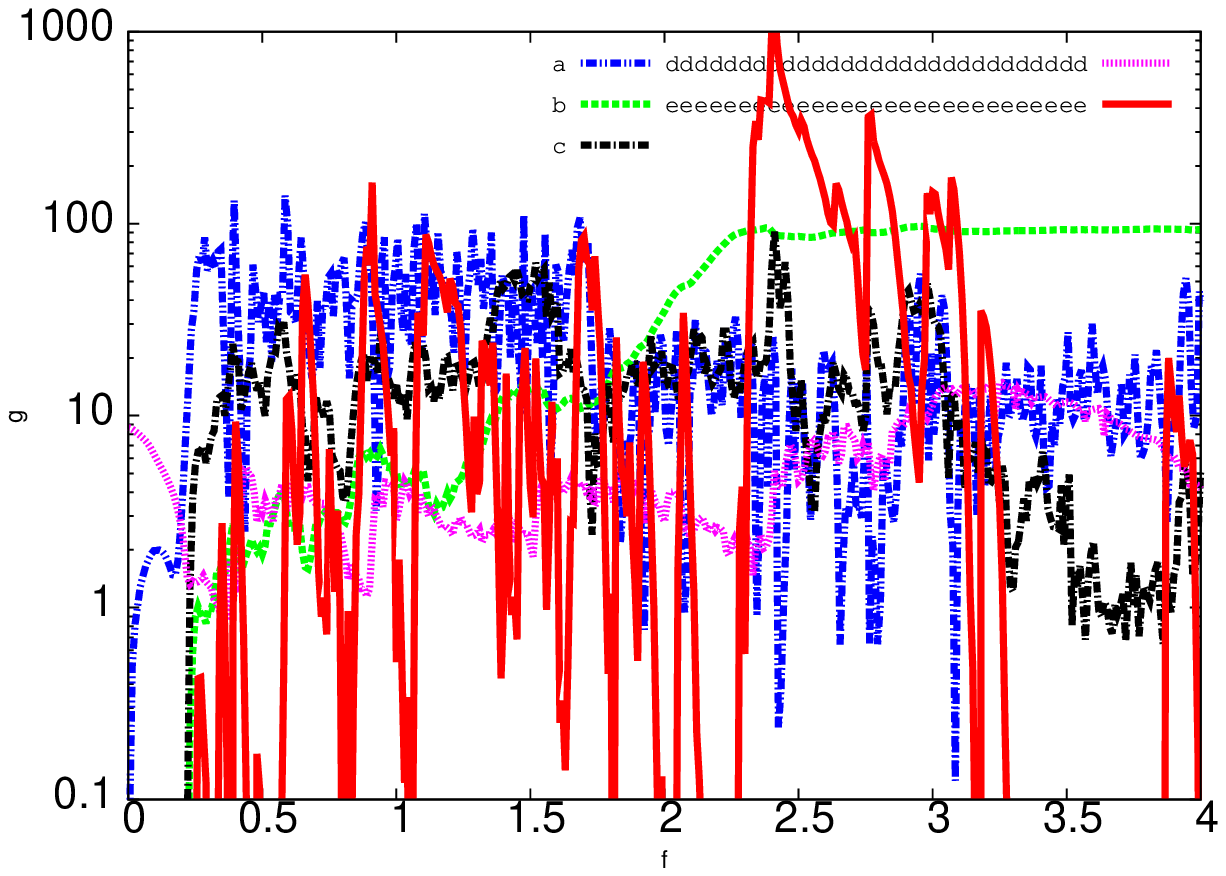}{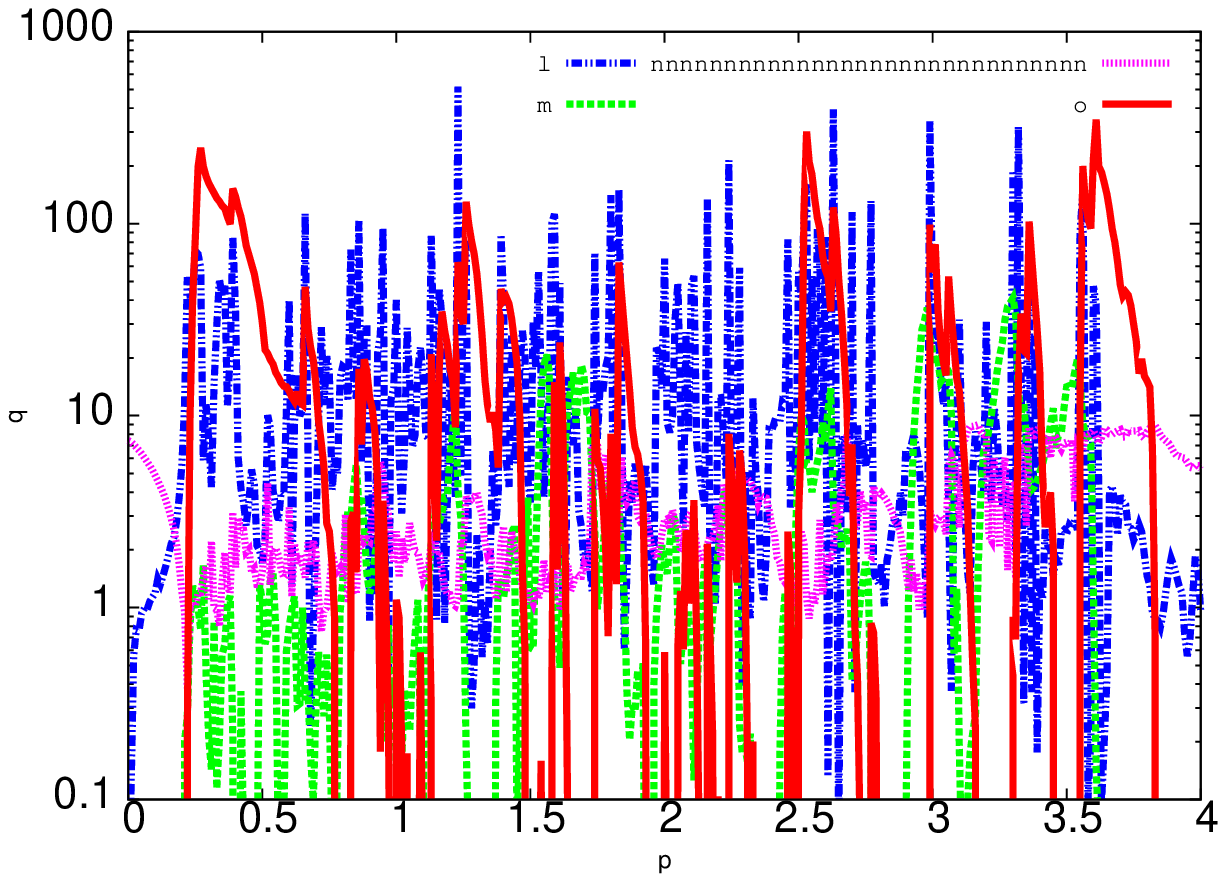} 
\caption{Various quantities (jet power, cold gas mass, radially-dominant cold gas mass, min[\tr], $\dot{M}_{\rm acc}$) 
as a function of time in the fiducial 3-D (top panel) and 2-D (bottom panel) cluster simulations. The data are sampled every 10 Myr. The cycle shown in Figure 
\ref{fig:3-D_cycle} are based on the top panel. All quantities, except total cold gas mass, are statistically similar in 3-D and 2-D. The total cold gas mass, which is dominated by 
the cold torus, is much larger in 3-D and builds up in time. However, the radially-dominant cold gas ($|v_r|>|v_\phi|$) mass in 3-D is similar to the total cold gas 
mass in 2-D.}
\label{fig:2-D3-D_vs_time}
\end{figure}

Figure \ref{fig:2-D3-D_vs_time} shows various important quantities, such as jet power, cold gas mass, mass accretion rate through the inner boundary, as a function
of time for the fiducial 3-D (upper panel) and 2-D (bottom panel) runs. Encouragingly, various quantities, except the total cold gas mass, show similar trends
with time in 2-D and 3-D. The total cold gas mass is much larger in 3-D because of the formation of a massive cold torus which is absent in 
axisymmetry. 

In both 2-D and 3-D runs min(\tr) varies in the range 1 to 10, and is roughly anti-correlated with $\dot{M}_{\rm acc}$ and jet power.
 The maximum jet power goes up to 
$\sim 10^{45}$ erg s$^{-1}$ in both cases. The mass accretion rate and hence feedback power injection (Eq. \ref{eq:fb}) is more spiky in 2-D (can go above 100 
$\msun$ yr$^{-1}$ for some times) because, unlike in 3-D, the cold gas that is accreted in 2-D covers full angle $2\pi$ in $\phi$ because of axisymmetry. 
The jet power, which is calculated by measuring the instantaneous jet thermal energy, depends on the average mass accretion rate over 
$\lesssim 100$ Myr rather than the instantaneous value. Another difference between 2-D and 3-D is
that cold gas can be totally removed (through the inner boundary) after strong feedback jet events in 2-D but this never happens in 3-D; cold gas 
(even the radially-dominant component)
is present at all times because it is very difficult to evaporate/accrete the massive rotating cold torus. There definitely is a depletion in the amount of
radially-dominant cold gas after a strong feedback event in 3-D (at $\sim 3.5$ Gyr in the top panel of Fig. \ref{fig:2-D3-D_vs_time}).

Although we have not explicitly shown jet power and cold gas `phase-space' plots for our 2-D fiducial run, we have verified that it shows cycles 
similar to the 3-D run (left \& right panels of Fig. \ref{fig:3-D_cycle}). Indeed, Figure \ref{fig:2-D3-D_vs_time} indicates that the 2-D runs should also 
show clock-wise cycles in
jet energy and cold gas mass as a function of min(\tr). These cycles just reflect the sudden rise in the accretion rate ($\dot{M}_{\rm acc}$) and jet power 
due to cold gas condensation and slow relaxation to equilibrium after overheating (notice the fast rise and slow decline in jet energy for individual 
jet events in both panels of Fig. \ref{fig:2-D3-D_vs_time}). 

\subsubsection{Dependence on jet efficiency \& halo mass}
\label{sec:eff}

\begin{figure}
\psfrag{a}[r][r][0.7][0]{cluster, $\epsilon=6\times 10^{-6}$}
\psfrag{b}[r][r][0.7][0]{cluster, $\epsilon=6 \times 10^{-5}$}
\psfrag{c}[r][r][0.7][0]{cluster, $\epsilon=5\times 10^{-4}$}
\psfrag{dddddddddddddddddddddddddddd}[r][r][0.7][0]{massive cluster, $\epsilon=6\times 10^{-5}$}
\psfrag{eeeeeeeeeeeeeeeeeeeeeeeeeeee}[r][r][0.7][0]{massive cluster, $\epsilon=5 \times 10^{-4}$}
\psfrag{g}[c][c][1][0]{$\dot{M}_{\rm acc} (M_{\odot}{\rm yr}^{-1})$}
\psfrag{f}[c][c][1][0]{time(Gyr)}
\epsscale{1.25}
\plotone{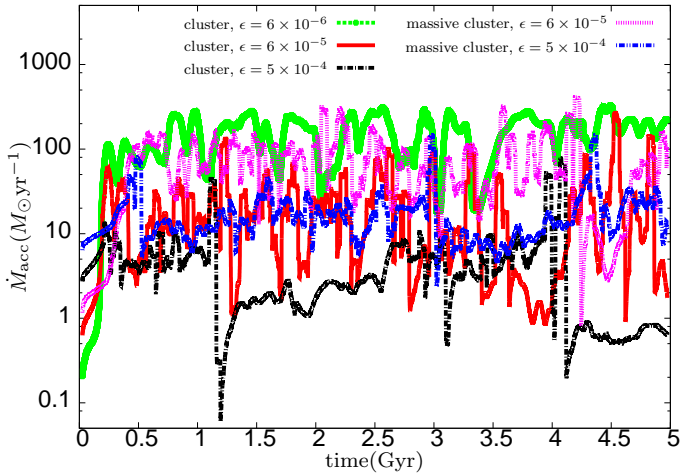}
\caption{The mass accretion rate through the inner radius (smoothed over 
50 Myr) as a function of time for different 2-D runs. A lower efficiency and a more massive halo lead to a larger accretion rate. }
\label{fig:mdot}
\end{figure}

Till now we have discussed the fiducial cluster simulation with a small feedback efficiency $\epsilon=6 \times 10^{-5}$. 
In this section we study the influence of jet efficiency ($\epsilon$) and halo mass ($M_{200}$) on various properties 
of the cluster core. Overall, we find that the effect of an increasing halo mass is similar to that of a decreasing feedback efficiency. We compare the efficiencies
($\epsilon$) ranging from $6 \times 10^{-6}$ to $0.01$. We consider two halo masses: a cluster with $M_{200}=7\times 10^{14} \msun$ and a massive cluster with
$M_{200}=1.8 \times 10^{15} \msun$.

Table \ref{table} and Figure \ref{fig:mdotbycf} show that the feedback efficiency
of $\epsilon=6\times 10^{-5}$ is able to suppress the cooling flow by about a factor of 10 for a cluster but only by a factor of 4 for a massive cluster.
This implies that a larger efficiency is required to suppress a cooling flow in a more 
massive halo. We note that the pure cooling flow accretion rate decreases with a decreasing halo mass because of a smaller amount of 
gas in lower mass halos (see the values enclosed in brackets in Table \ref{table}). 

Figure \ref{fig:mdotbycf} shows that the suppression factor ($\dot{M}_{\rm acc}/\dot{M}_{\rm cf}$) is smaller for the 
massive cluster, and scales as $\epsilon^{-2/3}$ for both cluster and massive cluster runs (see also Table \ref{table}). A decrease in the accretion rate with an 
increasing $\epsilon$ is not a surprise; a higher feedback efficiency heats 
the core more and maintains \tr$\gtrsim 10$ at most times, resulting in only a few cooling/feedback events. While the average jet power 
($\sim \epsilon \dot{M}_{\rm acc} c^2 \propto \epsilon^{1/3}$) increases with an increasing $\epsilon$, the core X-ray luminosity decreases. 
This implies that feedback heating and 
cooling do not balance each other at all times. Heating dominates cooling just after jet outbursts and cooling dominates in absence of 
infalling cold gas when \tr~ slowly decreases from a value $\gtrsim 10$. Thus, for a larger $\epsilon$, for which a cluster spends more time in a hot/dilute 
state, the X-ray emission from the core is expected to be smaller (c.f., Fig. \ref{fig:halomass_efficiency}).

Figure \ref{fig:mdot} shows the mass accretion rate (averaged over 50 Myr bins) as a function of time for our 2-D cluster and massive cluster runs with
different feedback efficiencies. The solid red line corresponds to the fiducial 2-D  cluster run (with 
$\epsilon=6\times 10^{-5}$). The green 
dotted line with a marker, which corresponds to a ten times lower efficiency ($\epsilon=6\times 10^{-6}$), shows an accretion rate comparable to a cooling flow
at most times (see also Table \ref{table}). The cluster run 
with ten times higher efficiency ($\epsilon=5 \times 10^{-4}$), indicated by black dot-dashed line, shows an average accretion rate of 5.1 $\msun~{\rm yr}^{-1}$ 
(about a fifth of the fiducial 2-D run; see Table \ref{table}); there are far fewer spikes in $\dot{M}_{\rm acc}$ compared to the fiducial run. Similar trends are 
observed for the massive cluster runs with $\epsilon=6\times 10^{-5}$ (magenta dotted line) and $5\times 10^{-4}$ (blue double-dot-dashed line). The number 
of  $\dot{M}$ spikes in Figure \ref{fig:mdot} are smaller for lower halo mass and higher feedback efficiency because of larger overheating and a longer recovery 
time after a precipitation-induced jet event.

\begin{figure*}
\psfrag{a}[r][r][0.8][0]{cluster, $\epsilon=6\times 10^{-5}$}
\psfrag{b}[r][r][0.8][0]{cluster, $\epsilon=10^{-4}$}
\psfrag{c}[r][r][0.8][0]{cluster, $\epsilon=5\times 10^{-4}$}
\psfrag{d}[r][r][0.8][0]{massive cluster, $\epsilon=6 \times 10^{-5}$}
\psfrag{e}[r][r][0.8][0]{massive cluster, $\epsilon=5 \times 10^{-4}$}
\psfrag{p}[c][c][1.2][0]{$r /r_{\rm max}$}
\psfrag{t}[c][c][1.2][0]{$T$(keV)}
\psfrag{s}[c][c][1.2][0]{$t_{\rm cool}/t_{\rm ff}$}
\psfrag{r}[c][c][1.2][0]{$n_e$(cm$^{-3}$)}
\psfrag{q}[c][c][1.2][0]{$ K (T_{\rm keV} / n_e^{2/3})$}
\epsscale{2.4}
\centering{
\includegraphics[width=3.5in]{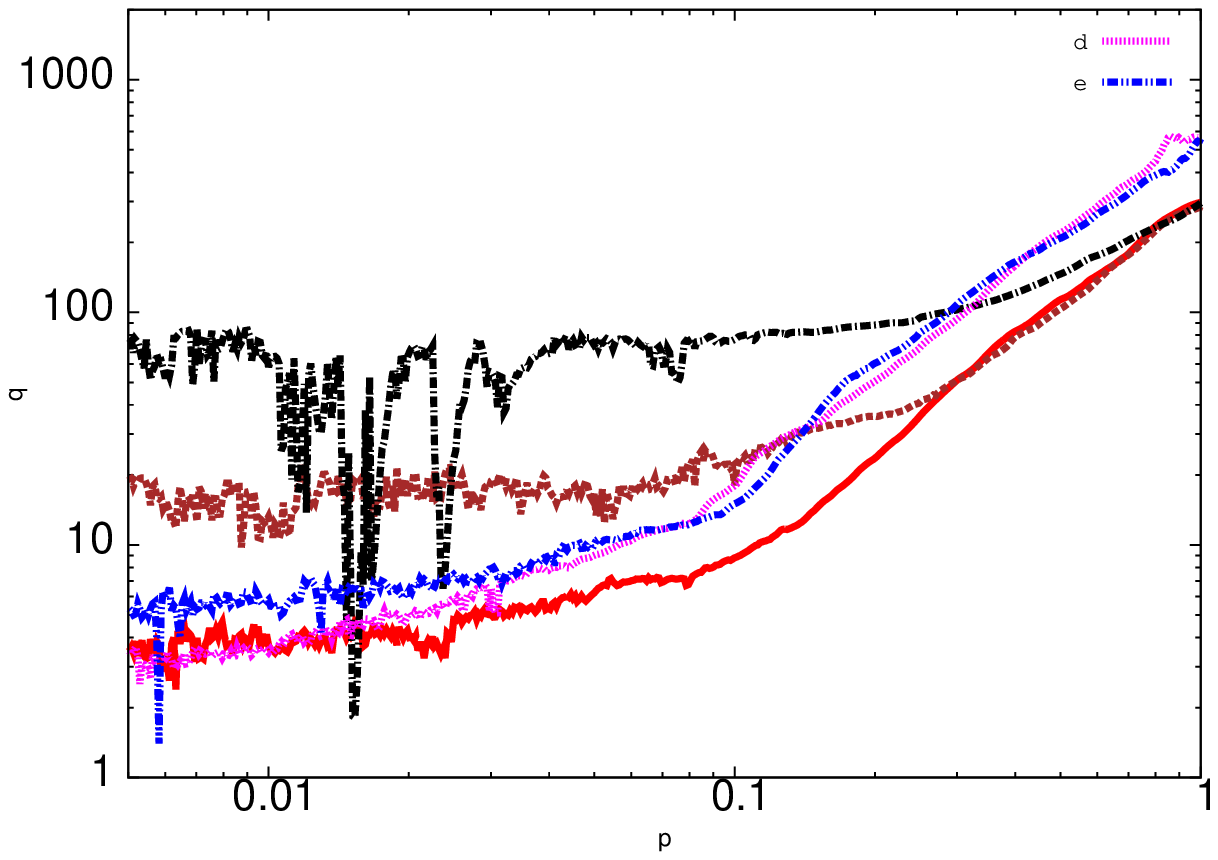}
\includegraphics[width=3.5in]{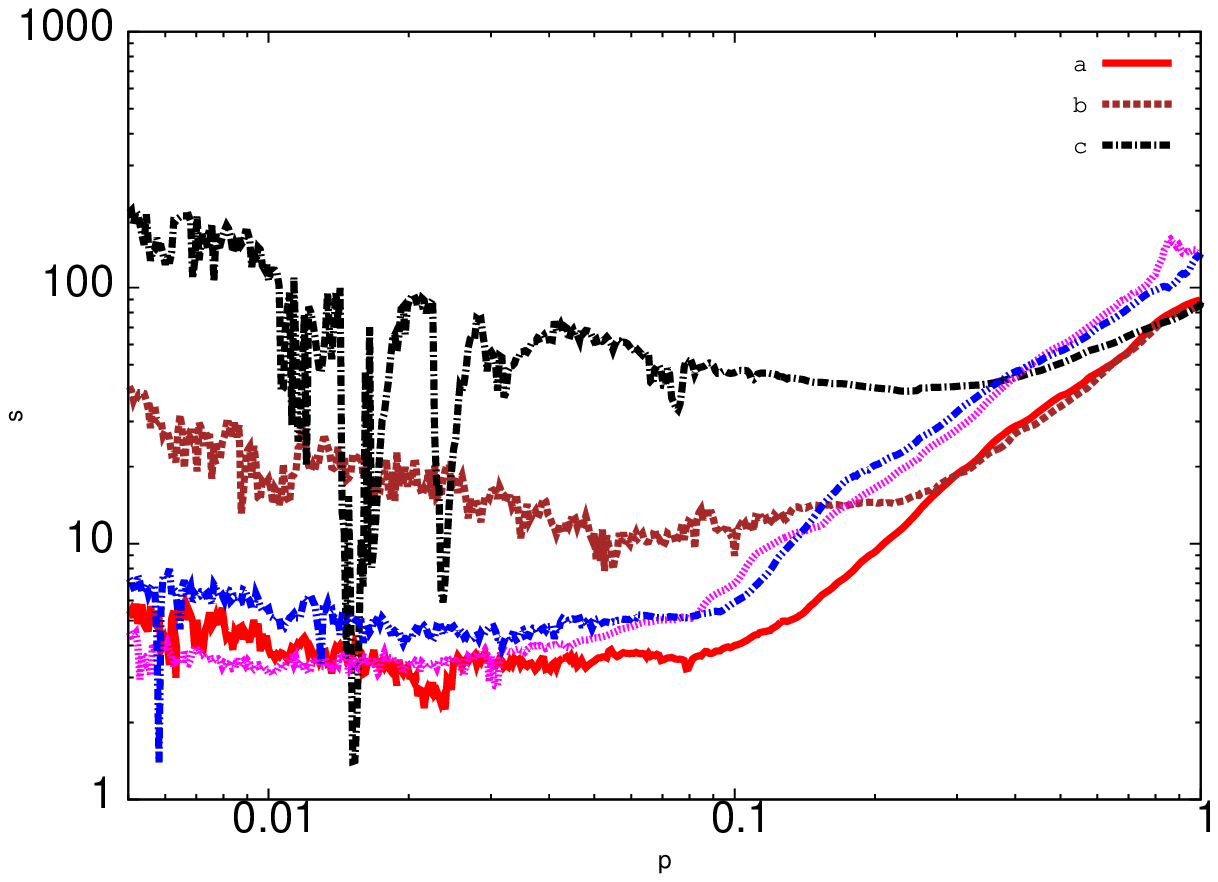} 
\includegraphics[width=3.5in]{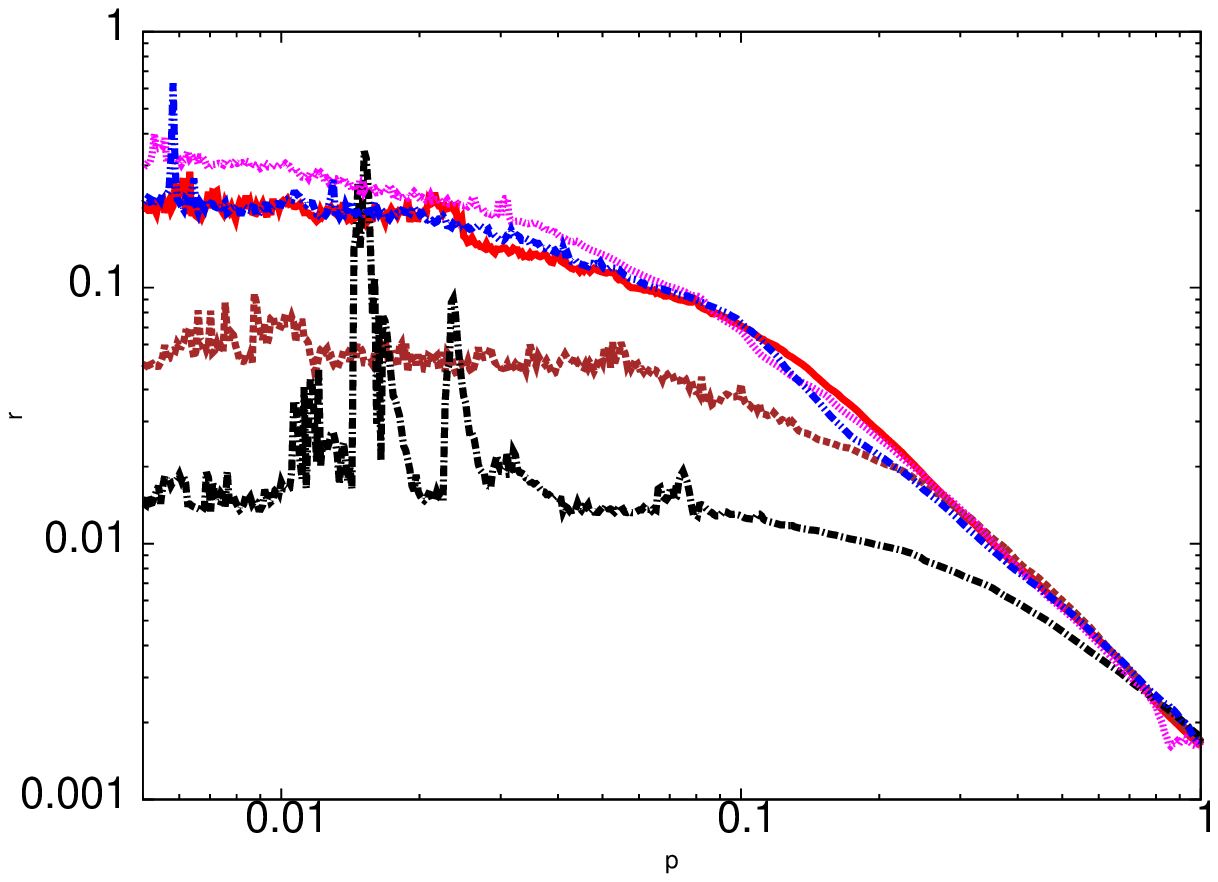}
\includegraphics[width=3.5in]{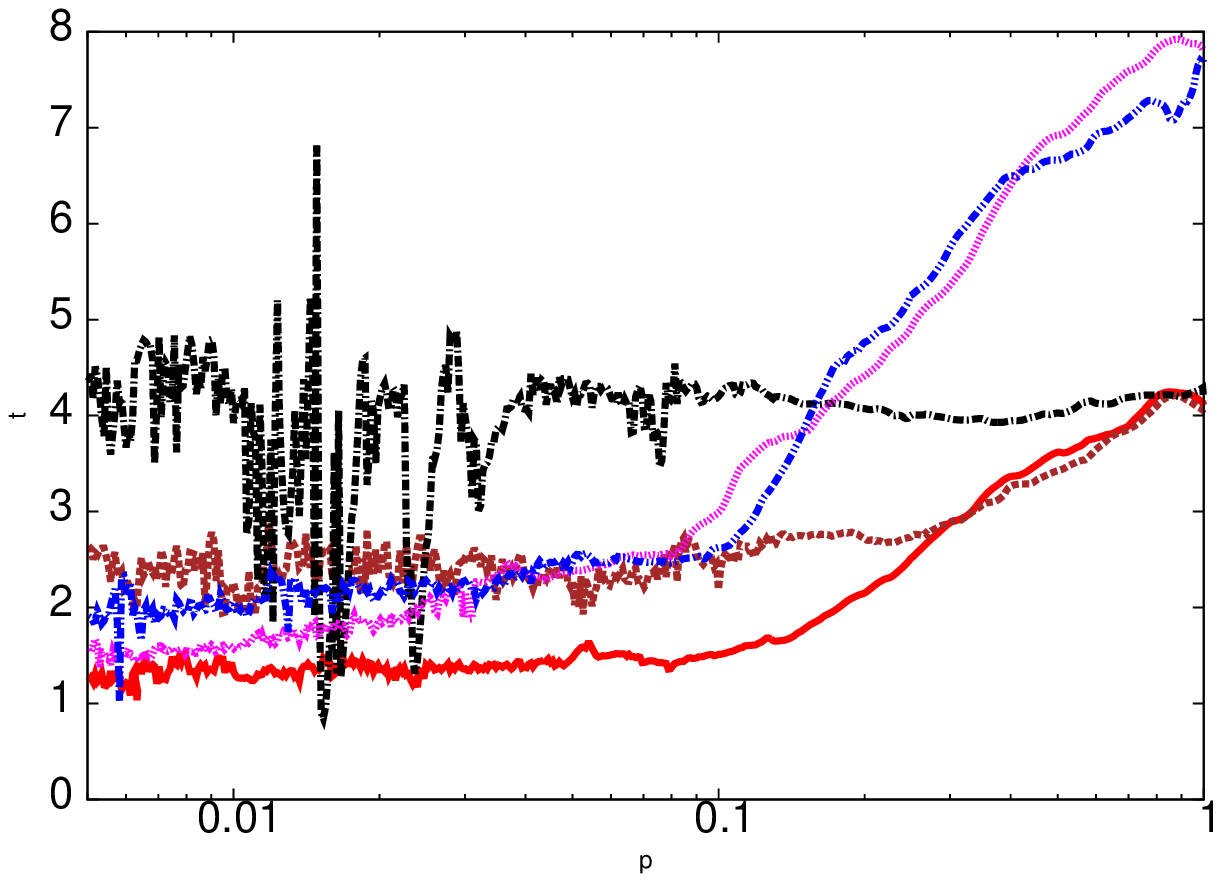}
}
\caption{Emissivity-weighted (considering plasma in the range $0.5-8$ keV), time (averaged between 4 and 5 Gyr) and angle-averaged profiles for 2-D 
runs as a function of radius scaled to the outer radius ($r_{\rm max}$): 
entropy ($K \equiv T_{\rm keV}/n_e^{2/3}$; top-left panel); $t_{\rm cool}/t_{\rm ff}$ (top-right panel); 
electron density ($n_e$; bottom-left panel); and temperature (in KeV; bottom-right panel). 
Both min(\tr) and core entropy decrease for a lower efficiency or  a larger halo mass. Temperature is higher for a higher efficiency, but a cool core (a 
temperature increasing with radius) is preserved for all cases, except the highest efficiency run. Spikes in the cluster run with $\epsilon=5\times 10^{-4}$
signify that there is cool, low entropy gas present in the core from 4 to 5 Gyr. 
}
\label{fig:entropy}
\end{figure*}

Figure \ref{fig:entropy} shows the time averaged (from 4 to 5 Gyr) and emissivity weighted (0.5-8 keV) 1-D profiles of several key quantities for 
2-D cluster and massive cluster runs with different efficiencies: entropy ($K \equiv T_{\rm keV}/n_e^{2/3}$), \tr~, number density and temperature. 
All profiles look similar to what is seen in observations. The entropy profile flattens
toward the center but the entropy core is prominent only for the higher efficiency ($\epsilon$) runs; a `core' with a constant \tr~is more prominent for 
lower $\epsilon$ and for the massive cluster. As expected, the density is lower and the temperature is higher for a larger feedback heating efficiency. For all 
efficiencies temperature increases with the radius (except for $\epsilon=5\times 10^{-4}$ which is almost isothermal just after a jet outburst; see the 
top-right panel of Fig. \ref{fig:halomass_efficiency}), as seen in the observations of cool-core clusters.

Compared to the cluster runs, the entropy for the massive cluster is higher at larger radii in Figure \ref{fig:entropy} because the initial entropy was scaled 
with the halo mass ($K \propto M_{200}^{2/3}$; see section \ref{sec:grid}). The entropy profiles for the massive cluster runs for the two 
efficiencies are similar; entropy keeps on decreasing as we go toward the center (forming a `core' in \tr), more so for $\epsilon=6\times 10^{-5}$. As we 
saw with the mass accretion rate in Figure \ref{fig:mdot}, the effect of increasing the efficiency is similar to that of decreasing the halo mass. {This is 
expected, as the mass accretion rate for lower mass halos is smaller, and the increase in jet efficiency and the consequent higher jet power 
suppresses accretion.\footnote{We 
thank the referee for the suggestion to highlight this point.}}

Another point to note in Figure \ref{fig:entropy} is that the profiles are rather similar for the massive cluster runs with $\epsilon=6\times 10^{-5}$ and 
$\epsilon=5\times 10^{-4}$. The bottom panels of Figure \ref{fig:halomass_efficiency} show that jet events between 4 to 5 Gyr are not able to raise 
min(\tr) much above 10 for these cases. However, top panels of Figure \ref{fig:halomass_efficiency}  and the bottom panel of Figure 
\ref{fig:2-D3-D_vs_time} show that between 4 to 5 Gyr min(\tr) increases with an increasing $\epsilon$. Therefore, the core entropy (density) for the cluster
runs increases (decreases) with an increasing $\epsilon$. Note that the core entropy for the cluster runs with a larger efficiency are not always higher; its 
only when the core is in the part of the heating cycle with min(\tr)$> 10$.

\begin{figure*}
\epsscale{1.25}
\plotone{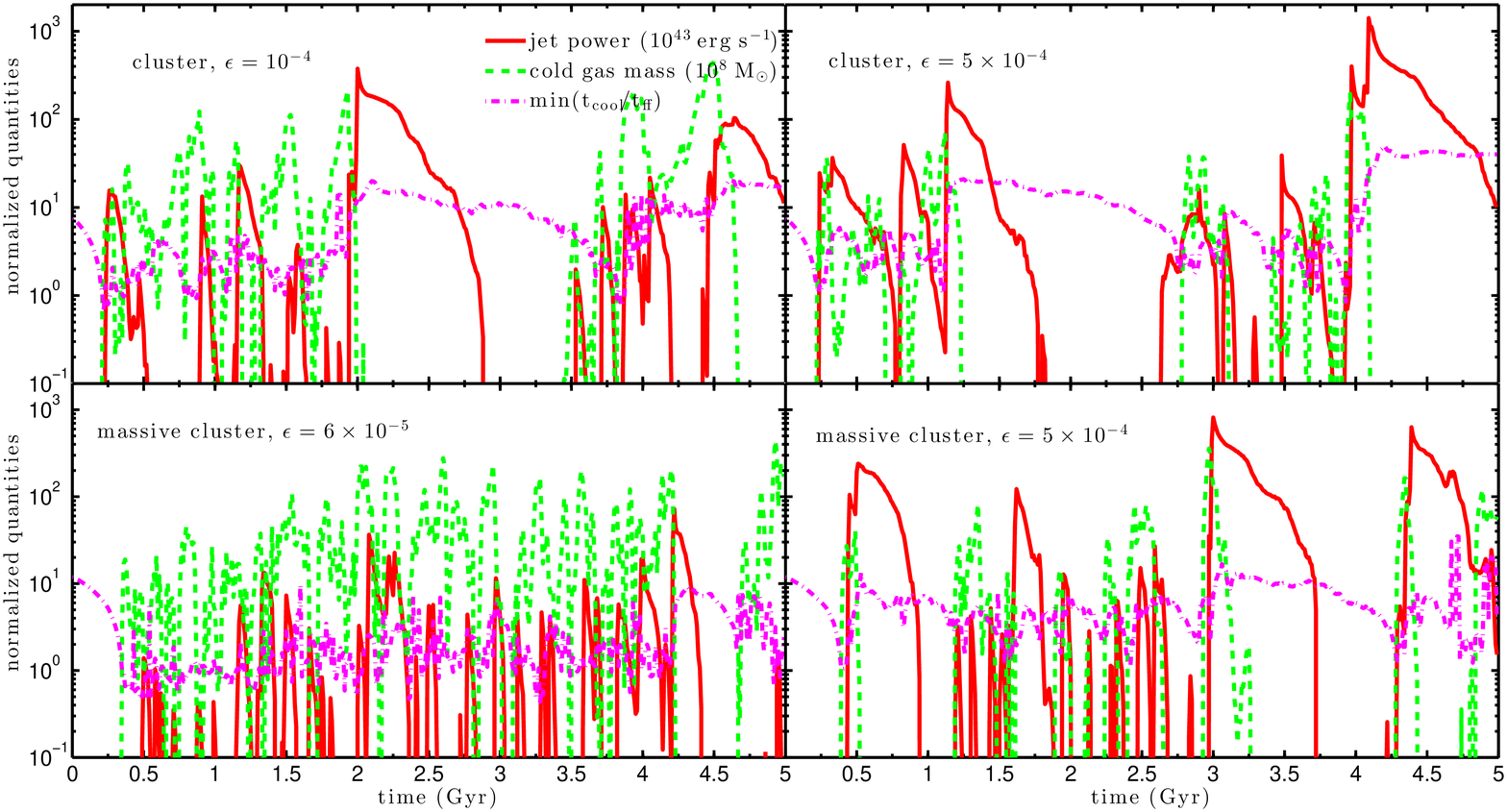}
\caption{Jet energy, cold gas mass, and min(\tr) as a function of time for 2-D runs with different efficiencies ($\epsilon=10^{-4},~5\times 10^{-4},~6\times 10^{-5}$)
and halo masses ($M_{200}=7\times 10^{14},~1.8\times10^{15} \msun$). Note that the jet energy and cold gas mass are scaled differently in different panels. 
A smaller efficiency or a larger halo mass leads to many accretion and jet feedback cycles.}
\label{fig:halomass_efficiency}
\end{figure*}

Figure \ref{fig:halomass_efficiency} shows various quantities (jet power, cold gas mass and min[\tr]) as a function of time for 2-D cluster 
($\epsilon=10^{-4}$ and $5\times 10^{-4}$; also see the 2-D cluster run with $\epsilon=6\times 10^{-5}$ in the bottom panel of Fig. \ref{fig:2-D3-D_vs_time}) 
and massive cluster ($\epsilon=6\times 10^{-5}$ and $5\times 10^{-4}$) runs. The first point to note is 
that the number of jet events (and hence the number of cycles; e.g., see Fig. 
\ref{fig:3-D_cycle}) is smaller for a higher efficiency and a lower halo mass. Another is that 
the peaks in jet power and min(\tr) for a higher efficiency are larger, resulting in overheating and longer durations for which cold gas and jet 
power are suppressed. Stronger overheating after jets in higher 
efficiency (and lower halo mass) runs results because, while the number of cold accretion events are smaller (compared to lower efficiency or a larger halo mass), 
the mass accretion rate during the multiphase cooling phase is similar (see Fig. \ref{fig:mdot}), generally giving larger heating (Eq. \ref{eq:fb}).

As with the mass accretion rate (see the spikes in Fig. \ref{fig:mdot}), for a fixed $\epsilon$ the number of cooling/jet
events are larger for a massive cluster. 
While \tr~ is $\lesssim 10$ and cold gas is present at most times for the massive cluster run with $\epsilon=6\times 10^{-5}$, there are longer periods 
with min[\tr]$\gtrsim 10$ and lack of cold gas for the cluster run (see bottom panel of Fig. \ref{fig:2-D3-D_vs_time}). The jet events are more disruptive (as measured 
by the rise in min[\tr]~after a jet event) in the lower mass halo because the jet power is relatively large but the hot gas mass is smaller (compare the right panels 
of Fig. \ref{fig:halomass_efficiency}).

\section{Discussion \& astrophysical implications}
\label{sec:dis}

The cold mode accretion model,\footnote{Here we use the label ``cold mode accretion" to refer to the capture and accretion of cold clouds by SMBH, and not the 
cosmological accretion of cold gas sometimes invoked in halos less massive than $10^{12} \msun$ (\citealt{bir03}).}  
in which local thermal instability leads to the condensation and precipitation of cold gas and enhanced accretion on to the SMBH, 
has emerged as a useful framework to interpret various properties in cores of elliptical galaxies, groups, and clusters (e.g., \citealt{piz05,sha12,gas12,li14b}). 
However, there are several unresolved problems: e.g., the role of angular momentum transport, self-gravity and cloud-cloud collisions in accretion on to the 
SMBH (e.g., \citealt{piz10,hob11,bab13,gas14}); relative contribution of cold gas at $\sim 1 $ kpc to SMBH accretion and star-formation; the role of thermal 
conduction in thermal balance and cold-gas precipitation (\citealt{wag14,voi14}); the exact mechanism 
(turbulent mixing, weak shocks; e.g., see \citealt{fab03,den05,ban14}) via which the mechanical power of the jet/cavity is dissipated and distributed throughout 
the core; and the scaling of various processes with the halo mass.

The key feature of cold mode accretion, unlike the hot mode, is that the mass accretion rate increases abruptly as \tr~becomes smaller than a critical value 
close to 10. This leads to a strong feedback heating, which temporarily overheats the cluster core. The hot mode feedback, in form of Bondi accretion onto 
the SMBH, on the other hand, is not an abrupt switch and increases smoothly with an increasing (decreasing) core density (temperature).
In section \ref{sec:comp} we discuss the success of the cold accretion model and compare with previous simulations. In section \ref{sec:comp_obs} we compare 
with the recent exquisite cold-gas observations and with statistical analyses of X-ray and radio observations of cluster cores.

\subsection{Comparison with previous simulations}
\label{sec:comp}

There are two broad categories of jet implementations described in the literature: first, where the jet mass, momentum and energy are injected via source terms 
(e.g., \citealt{omm04,cat07,li14,gas12}); second, where mass and energy are injected as flux through an inner boundary (e.g., \citealt{ver06,ste07}). 
We use the former approach, which has generally been more successful. 
 In this approach, the sudden injection of kinetic energy after cold gas precipitation leads to a shock which not only expands vertically but also laterally, 
 perpendicular to the direction of momentum injection. This lateral spread of jet energy and vorticity generation due to interaction with cold clumps help in 
 coupling the jet energy with the equatorial ICM. In the flux-driven approach the jet pressure is usually taken to be the same as ICM pressure and the jet drills 
 a cavity without expanding laterally in the core. Thus, coupling of the jet power is not very effective, unless the jet angle is very broad (\citealt{ste07}).

Our jet modeling is similar to the earliest works such as \citet{omm04,omm04b}, which inject jet mass, momentum and kinetic energy via source terms. However, this work 
focussed on the effect of a single jet outburst with a fixed power and did not include cooling; the simulations were run for short times ($\lesssim 500$ Myr). 
\citet{cat07} also implement jets using kinetic and thermal energy injection, and run for cosmological timescales. However, they use Bondi prescription
for accretion on to the SMBH, and hence their jet power input is tied to cooling and accretion at the center. 
Since the Bondi radius cannot be resolved in their simulations, they compute the Bondi accretion rate based on the density 
and temperature at a larger radius. Sometimes (e.g., in \citealt{dub10}) the Bondi accretion rate evaluated at large radii is artificially enhanced by a factor 
$\sim 100$ in order to match feedback heating with cooling. Bondi accretion is only applicable for a smooth, non-rotating gas 
distribution, and not for clumpy multiphase gas which can accrete at a much higher rate (e.g., \citealt{gas13,sha12}).

\citet{cie14} have studied the detailed structure and thermodynamics of source-term driven cylindrical jets, of different densities and temperatures, interacting 
with the ICM but they  run for less than 10 Myr. Like us, they also highlight the importance of hot back flows in regulating the central ICM.

Another set of simulations inflate cavities using jets driven by fluxes of mass and momentum at the inner radial boundary (rather than using source terms 
like us; in cluster context, see 
\citealt{ver06,ste07}; for MHD modeling of the Crab nebula jet, see \citealt{mig13}). \citet{ver06} injected momentum (and kinetic energy) via the inner 
radial boundary, with an opening angle of $15^0$, in form of 100 times hotter gas but in pressure equilibrium with the ICM. Their jets just drill through 
a narrow channel without coupling to the catastrophically cooling core.

 \citet{ste07} advocated wide (with opening angle $\gtrsim 50^0$) boundary-driven jets, such that the jet is not as fast, and can lead to vortices and substantial 
 mixing in cluster cores. However, since their simulations are not run for many cooling times, its unclear if wide jets and can indeed balance cooling for cosmological 
times. Moreover, the fat jets may not reproduce the observed morphologies of thin jets and fat bubbles. Using the boundary injection approach, \citet{hei06} 
emphasize the importance of the dynamic ICM in redistributing jet energy but they also run for less than a cooling time.

Recent numerical simulations of AGN-driven jets (\citealt{gas12,li14,li14b}) have been quite successful in producing several observed features such as, the 
lack of plasma cooling below a third of the ICM temperature (Fig. 11 in \citealt{li14b}), suppression of cooling and accretion in the core (by a factor of 10-100 
relative to a cooling flow), maintenance of cool-core structure even with strong intermittent jet events, formation of an angular momentum supported cold-gas 
torus, viability of AGN feedback from elliptical galaxies to massive clusters. Our simulations are different from these recent works, which use mesh refinement 
in a cartesian geometry, in that we use a spherical coordinate system. We have also tried to 
push the AGN feedback efficiency toward the lower limit which is still able to suppress a cooling flow. We find that an efficiency $\epsilon=6\times 10^{-5}$
is able to suppress a cluster cooling flow by a factor of 10. 

Like us, \citet{gas12} and \citet{li14b} also make an estimate of the mass accretion rate on to the SMBH. \citet{gas12} consider a spherical shell of radius 0.5 kpc
and calculate the mass accretion rate ($\dot{M}_{\rm acc}$) due to infalling gas. \citet{li14b,li15} calculate the mass accretion rate ($\dot{M}_{\rm acc}$) 
by dividing the cold gas mass within 0.5 kpc by 5 Myr (of order the dynamical timescale). Our estimate of $\dot{M}_{\rm acc}$ is similar to \citet{gas12}, except that
we calculate it at 1 kpc. Only a small fraction of $\dot{M}_{\rm acc}$ is expected to be accreted onto the SMBH; thus, the efficiency factor ($\epsilon$) in 
Eq. \ref{eq:fb} takes into account both the fraction of $\dot{M}_{\rm acc}$ that is accreted by the SMBH and the efficiency of converting SMBH accretion 
into jet mechanical energy.

While our jet feedback implementation is very similar to \citet{gas12}, our results differ in some key respects. The main difference is that 
we see extended cold gas and jet/cold-gas cycles even at late times (see Figs. \ref{fig:3-D_fiducial}, \ref{fig:3-D_torus}, 
\ref{fig:3-D_cycle}, \ref{fig:2-D3-D_vs_time}). 
Like \citet{li14b}, in \citet{gas12} there is a long-lived rotationally 
supported torus at few kpc and the 
extended multiphase gas is lacking at later times (see their Figs. 10 \& 11). The 
main reason for the absence of extended cold gas and strong jets at late times in previous simulations is a large feedback efficiency. 
A larger feedback efficiency leads to very strong feedback heating at early times, and the core reaches rough thermal balance in a state of \tr$>10$ with 
no fresh extended (radially dominant) cold gas condensing at late times. 
Since a large fraction of cool core clusters show extended cold gas (\citealt{mcd11}), a smaller value of 
feedback efficiency seems more consistent with observations.

\begin{figure*}
\epsscale{1.1}
\plotone{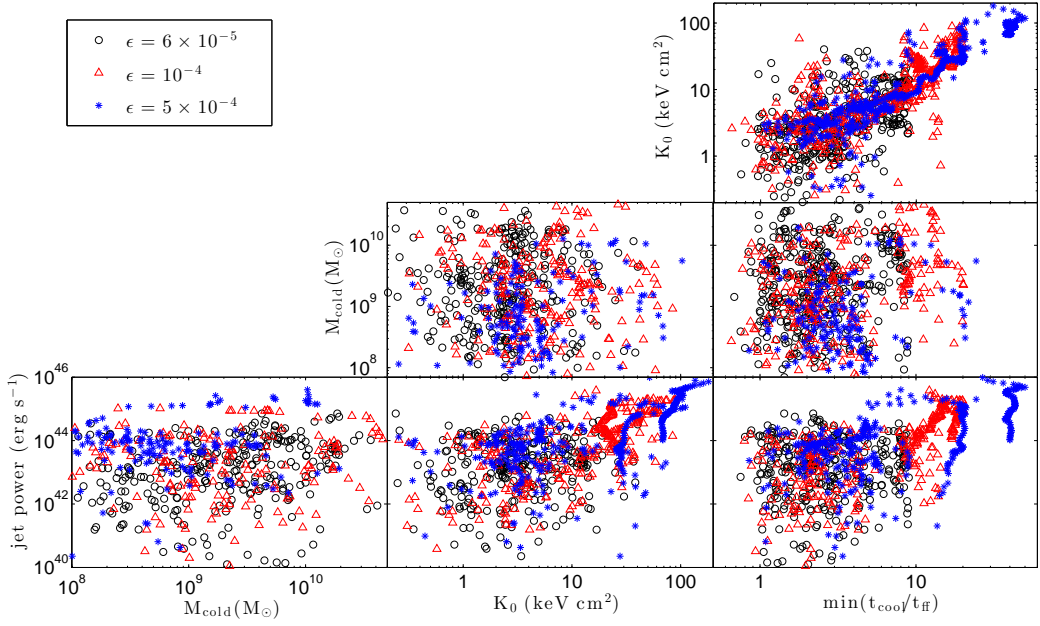}
\caption{Various important quantities measured at the same time  (jet energy, cold gas mass, core entropy, and min[\tr]) {plotted against each other} 
from our 2-D cluster 
runs with different efficiencies.  The data is sampled every 10 Myr. 
There is a strong correlation between the core entropy and min(\tr), especially at larger values of min(\tr). There is also a 
positive correlation between $K_0$-jet energy and min(\tr)-jet power. Larger efficiency runs lead to a larger value of min(\tr) and $K_0$. Notice that cold gas is 
absent if min(\tr)$\gtrsim 30$. }
\label{fig:corr_2-D}
\end{figure*}

\begin{figure*}
\epsscale{1.1}
\plotone{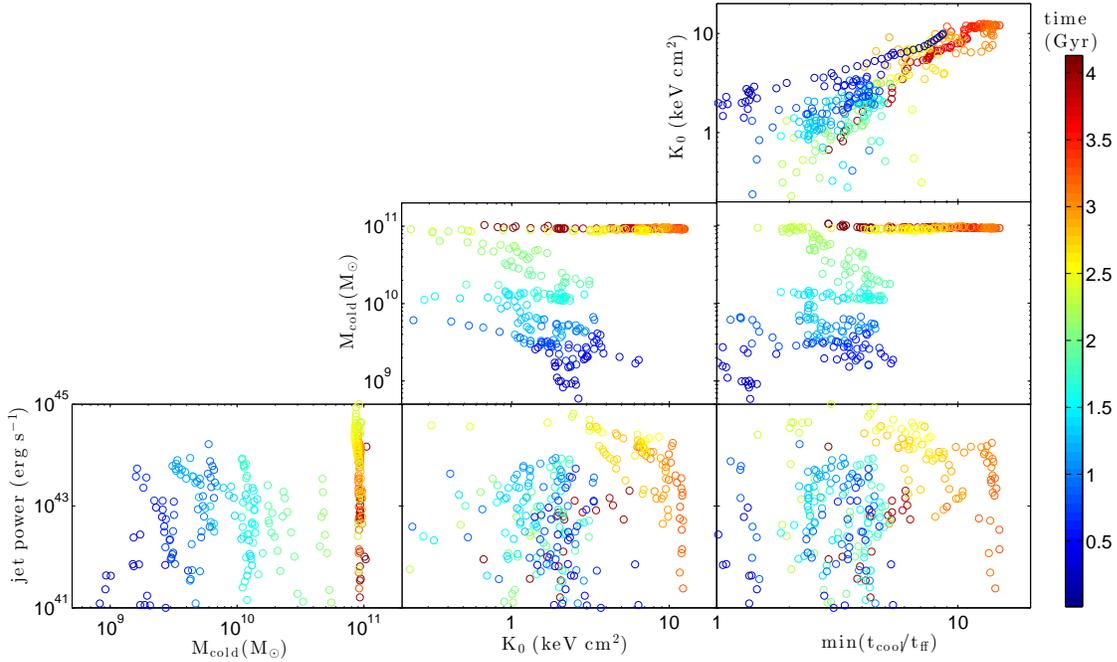}
\caption{Important quantities  measured at the same time  (jet energy, cold gas mass, core entropy, and min[\tr]) {plotted against each other} 
in our fiducial 3-D cluster 
run. As in 2-D runs (see Fig. \ref{fig:corr_2-D}), there is a strong correlation between $K_0$-min(\tr), $K_0-$jet power, and min(\tr)-jet energy. The cold gas mass
is high and becomes almost constant at later times as seen in the top panel of Fig. \ref{fig:2-D3-D_vs_time}. The color-coding corresponds to time. }
\label{fig:corr_3-D}
\end{figure*}


To solve the problem of a steady cold torus present at late times, very recently, \citet{li15} have incorporated the depletion of cold gas via star formation in 
the core, but they adopt the same feedback prescription as in their previous works. 
Star formation exhausts the amount of cold gas within 0.5 kpc, suppresses AGN heating, 
and leads to a cooling event after which cold gas condenses again. Thus, they obtain three cooling-feedback cycles in their fiducial run. In their AGN feedback 
prescription, star formation efficiency primarily determines the frequency of cooling/heating cycles. In contrast, our cycles are determined by the AGN 
feedback efficiency and the halo mass (more cycles for a massive halo and a smaller feedback efficiency). While there is some ongoing star formation in cool 
cluster cores, \citet{li15} form $3\times 10^{11} \msun$ in stars over 6.5 Gyr for their fiducial run corresponding to the Perseus cluster, a significant fraction of 
the mass of the BCG (brightest cluster galaxy; $8\times 10^{11} \msun$; \citealt{lim08}). This does not agree with semi-analytic models which suggest that 80\% of the stars of BCGs 
are assembled before $z=3$ (i.e., only $2\times 10^{10} \msun$ are expected to form over the past 12 Gyr; \citealt{del07}). Moreover, the current star formation 
rate even in the most extreme cool core clusters is typically $\lesssim 10 \msun~{\rm yr}^{-1}$ (\citealt{hic10,mcd11b}); the average star formation rate 
of \citet{li15} would correspond to an unacceptably high value, $\approx 46 \msun~{\rm yr}^{-1}$.

We can directly compare our results with \citet{gas12} as our feedback prescription is similar. They tried jet efficiency factors of $\epsilon=6\times 
10^{-3},~0.01$. With a much higher accretion efficiency ($\epsilon=0.01$) compared to ours ($\sim 10^{-4}-10^{-5}$), 
\citet{gas12} get a larger suppression factor ($\dot{M}_{\rm acc}/\dot{M}_{\rm cf}\sim 1-2\%$) compared to us, as expected. \citet{gas12} use a halo mass 
$M_{200} \sim 10^{15} \msun$. Figure \ref{fig:mdotbycf} shows how our results compare with \citet{gas12}. 
The suppression factor of a massive cluster ($M_{200}=1.7 \times 10^{15} \msun$) for our fiducial $\epsilon=6\times 10^{-5}$ is 20\%, 
larger than their work. Suppression factor in our massive cluster (cluster) run for $\epsilon=0.01$, as seen in Figure \ref{fig:mdotbycf}, is 3\% (0.8\%), 
in rough agreement with the results of \citet{gas12}.

\subsection{Comparison with observations}
\label{sec:comp_obs}

Now that we have done some comparisons with previous simulations, in this section we compare our results with observations. We note that the observational
comparison may not be perfect because our simulations lack some physical processes such as magnetic fields and thermal conduction. These effects will be 
considered later. Moreover, observations suffer from projections effects, and our cluster parameters do not span as broad a range 
(of halo masses, entropy profiles at large radii, etc.) as  encountered in observations.

One of the most commonly studied ICM property is its entropy profile (e.g., \citealt{cav09}). Figure \ref{fig:entropy}
shows the time averaged (4-5 Gyr), X-ray emissivity weighted profiles of \tr~ and entropy as a function of radius for our various runs. We see that an 
entropy core (with a prominent local minimum in \tr) is a good approximation
for systems in which \tr$\gtrsim 10$ and there is no extended cold gas. This state is common for simulations with larger efficiency and lower halo mass. 
However, the halos with \tr$<10$, in which there is lot of currently condensing and infalling cold gas, are consistent with a `core' in \tr~and a decreasing entropy toward
the cluster center, albeit with a shallower slope. This is consistent with recent reanalysis of core entropy profiles (\citealt{pan14}), which suggests that a 
double power-law entropy profile, with a shallower entropy in the center, better describes the ICM core. It will be useful to compare the behavior of central 
entropy as a function of min(\tr); we expect entropy cores for  min(\tr)$>10$ and slowly increasing entropy profiles for min(\tr)$\lesssim 5$.

Figures \ref{fig:corr_2-D} and \ref{fig:corr_3-D}  show the correlation between various important quantities
 for our 2-D cluster runs (with $\epsilon=6\times 10^{-5},~10^{-4},~5\times 10^{-4}$) and the 3-D fiducial run, respectively. Data points sampled 
 every 10 Myr are shown. The core entropy ($K_0$) is obtained by using
a least squares fit to the emissivity-weighted 1-D entropy profile of gas in 0.5-8 keV range. In both these figures the strongest correlation is 
between $K_0$ and min(\tr), as expected,
because both these quantities depend on density and temperature in a similar way ($K \propto T/n^{2/3}$ and \tr$\propto T^{1/2}/n$; see Eq. 35 
in \citealt{mcc12}); the relation is not one-to-one because $K_0$ is determined by entropy near the center and min(\tr) by the behavior at the core radius 
(beyond which density decreases sharply). 

The spread in $K_0-{\rm min(}$\tr) correlation is larger for a lower $K_0$ (or equivalently, min[\tr]; this is also 
seen in observational data shown in Fig. 4 in \citealt{voi14}) because a core 
with constant entropy is not a good description in that case and the entropy decreases inward (see top-left panel of Fig. \ref{fig:entropy}).

\begin{figure*}
\epsscale{1.2}
\plotone{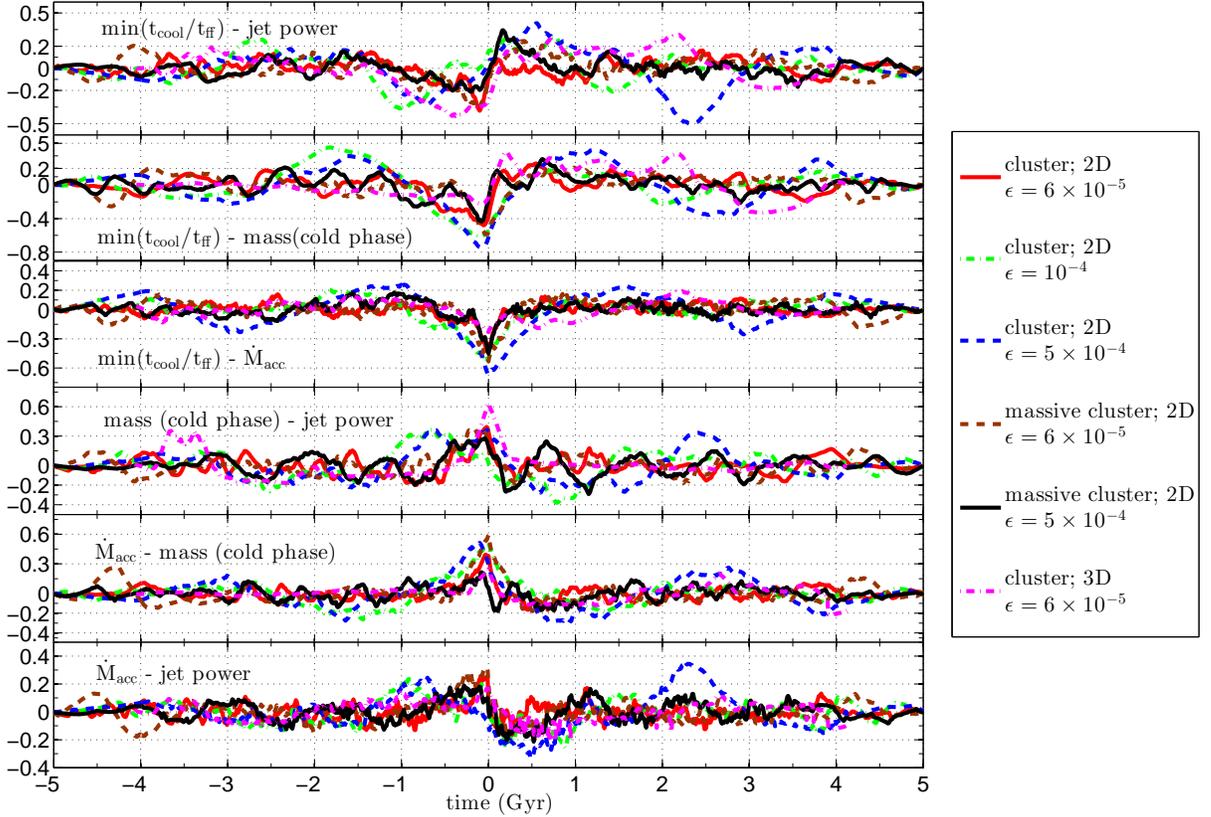}
\caption{Cross-covariance of various quantities (min[\tr], jet energy, mass in cold phase, $\dot{M}_{\rm acc}$) as a function of time lag to show temporal relationship 
between these various quantities. Cross covariance between two quantities as a function of time, as used here, is defined as: 
cov($a,b;\tau$) = $\int_0^{T-|\tau|} \left [ \delta a (t+\tau) \delta b (t) dt \right ]/ \left [ \sqrt{ \int_0^T |\delta a(t)|^2 dt  \int_0^T |\delta b(t)|^2 dt } \right ] $, 
where $-T \leq \tau \leq T$ is the time lag and $\delta a$ and $\delta b$ are mean-subtracted quantities. Since there is a large variation in various quantities (see 
Figs. \ref{fig:corr_2-D} \& \ref{fig:corr_3-D}), we take log before evaluating cross-covariance. For the 3-D cluster run we have used 
the radially-dominant cold gas mass; the cross-covariance is much weaker if we use total cold gas mass.}
\label{fig:corr_t}
\end{figure*}

The correlation between various quantities in Figures \ref{fig:corr_2-D} and \ref{fig:corr_3-D} are not  particularly strong because of the hysteresis 
behavior of various quantities (e.g., jet power, radially dominant cold gas mass) with respect to the core properties (Fig. \ref{fig:3-D_cycle}). 
Figures \ref{fig:corr_2-D} \& \ref{fig:corr_3-D} show that, in general, the jet power increases for a larger $K_0$ (or min[\tr]), particularly for a larger core entropy. 
This is because a large jet power
overheats the cluster core and raises its entropy. Other quantities do not show as strong correlations in these plots; cold gas mass increases with a lower
entropy or a shorter cooling time, but jet energy and cold gas mass show a large spread relative to each other (since cold gas leads to increase in jet energy, 
which in turn suppresses cold gas mass). 

The 3-D run shown in Figure \ref{fig:corr_3-D} prominently shows the sign of the massive torus at late times. Apart 
from this, there are no major differences in 2-D and 3-D. Also note that cold gas is missing in $\epsilon=5\times10^{-4}$ 2-D cluster run for \tr$\gtrsim 20$ 
(Fig. \ref{fig:corr_2-D}; the same is expected for the radially dominant cold gas in 3-D). This is consistent with the observations of \citet{cav08}, 
who find that H$\alpha$ luminosity
is suppressed for a core entropy $>$ 30 keV cm$^2$ (corresponding to min[\tr] of about 20; see the top-right panel of Fig. \ref{fig:corr_2-D} \&
 Fig. 4 in \citealt{voi14}). The onset of 
star formation in cluster cores also happens sharply below the same entropy threshold (\citealt{raf08}). Figure \ref{fig:corr_3-D} shows that the core entropy
and min(\tr) remain below 20 even if the instantaneous jet power is as high as $10^{45}$ erg s$^{-1}$; core entropy can be much higher (up to 100 keV cm$^2$) 
for higher efficiency (see $\epsilon=5 \times 10^{-4}$ in Fig. \ref{fig:corr_2-D}).

While Figures \ref{fig:corr_2-D} \& \ref{fig:corr_3-D} show correlations between various quantities at a given time, we are also interested in understanding 
causal relationships between various quantities such as cold gas mass, min(\tr), jet energy.  Figure \ref{fig:corr_t} shows the temporal cross-covariance 
between various important parameters (we take log before calculating cross-covariance). Various 2-D and 3-D simulations with different efficiencies are 
plotted together. The trends which are common to all simulations are likely to be robust. Robust 
correlations among various quantities occur only with a time lag $< 1 $ Gyr (typical core cooling/heating time). 

Top three panels in Figure \ref{fig:corr_t} show that the cross-covariance between min(\tr) decreases $\sim 0.5$ Gyr before zero lag and rises for 0.5 Gyr 
after that (rise is not so prominent with $\dot{M}_{\rm acc}$). The interpretation is that the cooling time (and hence min[\tr]) decreases as the core cools
during the cooling leg of the cycle. This leads to the condensation of radially dominant cold gas after a cooling time (few 100 Myr) and an enhancement of 
the mass accretion rate and the jet power. Sudden increase in jet power overheats the core and min(\tr) increases after a lag of few 100 Myr; cold gas mass 
and $\dot{M}_{\rm acc}$ also decrease consequently.

The bottom three panels in Figure \ref{fig:corr_t} show that cold gas mass (radially dominant), mass accretion rate, and jet power are positively correlated. A 
slight skew toward a negative time lag shows that the cold gas mass and the mass accretion rate increases first, and that gives rise to an increase in jet power.
Thus, the cross-covariance behavior of different variables is similar to that seen in cold gas-jet cycles in Figure \ref{fig:3-D_cycle}. Also note in Figure 
\ref{fig:corr_t}  that there are smaller number of oscillations for higher feedback efficiency (or smaller halo mass); this is a reflection of smaller number of
cooling/feedback cycles in these cases (see Fig. \ref{fig:halomass_efficiency}).

In our 3-D simulations we see the build-up of a massive rotationally-supported torus. A part of this torus should cool further and lead to star formation, 
as argued by \citet{li15}. While the cold gas (mostly in the torus) builds up in time and saturates after 2 Gyr, the jet energy shows fluctuations in time 
even after that (see the top panel of Fig. \ref{fig:2-D3-D_vs_time} and Fig. \ref{fig:corr_3-D}).
Therefore, in our models there is no correlation between {\em total} cold gas mass and jet energy. However, there is a correlation between the radially-dominant 
cold gas mass and the jet energy (compare green-dotted and black dot-dashed lines in  the top panel of Fig. \ref{fig:2-D3-D_vs_time}). Thus, 
although most cold gas is decoupled 
from jet feedback, it is the subdominant {\em in-falling cold gas} which is powering AGN. This is in line with the observations of \citet{mcn11}, who find 
no correlation between the jet power and the available molecular gas. They, therefore, argue that most of the cold gas is converted into stars
rather than being accreted by the SMBH.

Finally, we compare our simulations with recent observational studies of cold gas kinematics and star formation. These have been studied in 
unprecedented detail in some elliptical galaxies and clusters, thanks mainly to 
{\em ALMA} and {\em Herschel} telescopes (e.g., \citealt{mcn14,rus14,dav14,edg10,wer14,tre12,raw12}). 
In this paper we have mainly focussed time-averaged kinematics, as shown in Figures \ref{fig:vel_r_phase} and \ref{fig:velpdf}. We can clearly see three
kinematically distinct components of cold gas: a rotationally-supported massive torus, ballistically infalling cold gas, and jet-uplifted fast cold gas.

Observations of different clusters are snapshots at a particular instant, 
at which a particular component (e.g., the rotating torus, a fast outflow, or a radially distributed inflow; see Fig. \ref{fig:jet_in_out}) of the 
cold gas distribution may be more prominent.
We will present the details of cold gas kinematics in various states of the ICM (with cold inflows, outflows, and the rotating torus) in a future work.

Some of the salient properties of the cold gas distribution in the fiducial run are: the rotating cold gas torus, when present, is more massive compared 
to in-falling cold gas (this 
component may be exaggerated in our simulations as we do not include star formation that would quickly consume some of the cold gas); the 
rotating disk rotates at the 
almost constant local circular velocity (100-200 km s$^{-1}$ for our fiducial cluster run; the actual value may be larger because we have ignored the 
gravitational potential due to the BCG) in form of a massive torus within 5 kpc; the radially-dominant cold gas is much more spatially extended (out to few 
10s of kpc) compared to the rotating torus, and the majority of this component also has a velocity close to the circular velocity; some 
(about $10^5 \msun$) radially dominant outflowing gas  has a radial velocity as high as  1000 km s$^{-1}$ (a fast component is seen in the observations 
of \citealt{rus14} and \citealt{mcn14}); the in-falling cold gas (on average) is about twice as much as the 
outflowing component. 

While the accretion rate through the inner radius (dominated by cold gas) is smaller than $100 \msun~{\rm yr}^{-1}$ (the accretion 
rate on to the SMBH is $\lesssim 1\%$ of this) at all times (see Fig. \ref{fig:2-D3-D_vs_time}), the cooling/accretion and outflow rates in the cold gas 
can be much larger instantaneously because of the massive cold torus buffer (Fig. \ref{fig:jet_in_out}). 

The observations show varying cold gas kinematics in different systems: radially in-falling cold molecular clouds of $3\times10^5$ to 
$10^7 \msun$ in a galaxy group NGC 5044 (\citealt{dav14}); $5\times 10^{10} \msun$ molecular gas predominantly in a rotating disk, and 
about $10^{10} \msun$ in a fast (line of sight velocity up to 500 km s$^{-1}$) outflow in Abell 1835 (\citealt{mcn14}); $10^{10} \msun$ of molecular
gas roughly equally divided between as rotating disk (velocity $\sim 250$ km s$^{-1}$) and a faster (570 km s$^{-1}$) infalling/outflowing component 
in Abell 1664 (\citealt{rus14}). 
In our simulations we observe similar components of the cold gas distribution, as shown in Figures \ref{fig:vel_r_phase} and \ref{fig:velpdf}.

\section{Conclusions}
\label{sec:con}
Cold-mode feedback, due to condensation of cold gas from the hot ICM when the local density is higher than a critical value (see below), 
has emerged as an attractive paradigm to interpret observations in cluster cool cores. In this paper we have carried out simulations of clusters of
{halo masses $7\times 10^{14} \msun$ and $1.8\times 10^{15} \msun$} with feedback driven AGN jets, {varying the feedback efficiency over a large 
range ($10^{-5}-0.01$)}. 
AGN feedback is able to suppress cooling flows within the observational limits ({by a factor of $\sim 10$}) {even for a 
feedback efficiency as low as $6 \times 10^{-5}$. This is the major difference from previous jet simulations, which use a much larger feedback 
efficiency ($\gtrsim 10^{-3}$; \citealt{gas12,li14b,li15}). Because of the high feedback efficiency, the previous simulations attain thermal equilibrium in
a hot, low-density core (with \tr$>10$) which does not show cold gas and jet cycles at late times. In contrast, our low efficiency simulations show cooling/jet 
cycles even at late times.}

The core undergoes cooling and feedback heating cycles because of cold gas precipitation and enhanced accretion on to the SMBH. {There are more cycles 
for a lower efficiency and a larger halo mass.}
The cool-core appearance is preserved even during strong jet events. {Even with large efficiencies, jet feedback raises the core entropy 
to several tens of keV cm$^2$, and therefore cannot explain the non-cool-core clusters with large cores and entropies greater than 100 keV cm$^2$. 
The origin of these non-cool-core clusters is still poorly understood (see, e.g., \citealt{poo08}).}

{In this paper}  we highlight some results that were not
emphasized in previous simulations of AGN jet feedback in clusters; in particular, we compare our results with several recent observations. 
Following are our major conclusions:

\begin{itemize}

\item First and most importantly, the results from different codes, different setups, and different implementation of jet feedback (as long as 
condensation and accretion of cold gas is accounted for; e.g., \citealt{gas12,li15}) give qualitatively similar results. This indicates the robustness of the 
cold feedback mechanism, and the importance of precipitation (which occurs when \tr$\lesssim 10$) and associated feedback in 
regulating cluster cores.

\item We find that a feedback efficiency (defined as the ratio of jet mechanical luminosity and the rest mass accretion rate [$\dot{M}_{\rm acc} c^2$] 
at $\sim 1$ kpc;  see
Eq. \ref{eq:fb}) of as small as $6 \times 10^{-5}$ is sufficient to suppress the cooling/star-formation rate in  cluster cores by a factor of about 10 (see Fig. 
\ref{fig:mdotbycf}). 
An even smaller efficiency is sufficient for lower mass halos because the thermal energy of the ICM is smaller compared to the rest mass energy. 
{Our fiducial efficiency
is at least 20 times lower than the models of \citet{li14b} and \citet{gas12}. Our values are consistent with the expectation that the mass accretion rate on to 
the SMBH is much smaller than the accretion rate estimated at $\sim 1$ kpc, and the fact that powerful jets exist only when the SMBH accretion rate 
is smaller than 0.01 time the Eddington rate.}

The required efficiency can be roughly estimated as follows. On average, the core luminosity 
is balanced by the energy input rate; i.e., 
$ \epsilon_{\rm req} \dot{M}_{\rm acc} c^2 \sim L_X $
($\epsilon_{\rm req}$ is the required feedback efficiency, 
$\dot{M}_{\rm acc}$ is the accretion rate estimated at 1 kpc, and $L_X$ is the X-ray luminosity of the cooling 
core). If we assume that the mass accretion rate is a fixed factor $f~(\ll 1)$ of the cooling flow value then, 
$$ 
\epsilon_{\rm req} f \frac{M_c c^2}{t_{\rm cool}} \approx \frac{1.5 k_B T M_c}{\mu m_p t_{\rm cool}}
$$ 
($M_c$ is the core mass) implies that 
$$ 
\epsilon_{\rm req} \approx \frac{c_s^2}{fc^2} \approx 3 \times 10^{-5} \left ( \frac{f}{0.1} \right )^{-1} \left ( \frac{M_{200}}{7 \times 10^{14} \msun} \right)^{2/3},
$$ 
normalizing to the parameters of our fiducial cluster ($T=2$ keV; see the bottom right panel of Fig. \ref{fig:entropy}), where $c_s^2 = k_BT/\mu m_p$ is 
the sound speed of the core ICM. {This estimate agrees with our fiducial efficiency  $\epsilon = 6 \times 10^{-5}$.}


\item We observe cycles in jet energy, radially-dominant cold gas mass, and mass accretion 
rate which are governed by \tr~ measured in the hot phase. If \tr$\lesssim 10$ cold gas precipitates, and leads to multiphase cooling and enhanced accretion on to
the SMBH. Sudden rise in the accretion rate, for a sufficiently high feedback efficiency, leads to overheating of the core and an increase in \tr~ above 
the threshold for cold gas condensation. We emphasize that thermal equilibrium in cluster cores only holds in a time-averaged sense.
There are cooling/heating cycles during which the core slowly cools/heats up. The core spends a longer time in the hot state for a larger feedback 
efficiency and a lower halo mass, leading to a smaller number of cooling/heating cycles (see Fig. \ref{fig:halomass_efficiency}).

Several observations hint at cycles in jet power and cooling of the hot gas (see Fig. \ref{fig:3-D_cycle}). 
We do not expect such cycles if feedback occurs via the smooth hot/Bondi mode as we do not have sudden cooling/feedback events.
The hysteresis behavior observed in the core X-ray properties ($K_0$, min[\tr]) and the mass of cold gas, jet power, etc. leads to a large dispersion in the 
correlation between these quantities (see Figs. \ref{fig:corr_2-D} \& \ref{fig:corr_3-D}). In particular, the mass accretion rate (at 1 kpc) is independent of the {\em total} 
cold gas mass, which is dominated by the rotating cold torus, most of which is consumed by star formation (rather than accretion on to SMBH; e.g., \citealt{mcn11}).

\item We can classify the cold gas in our 3-D simulations into two spatially and kinematically distinct components: a centrally concentrated (within 5 kpc), 
rotationally supported ($|v_\phi| \gg |v_r|$) torus (Fig. \ref{fig:3-D_torus}); and extended (both infalling and outgoing) cold gas going out to 30 kpc 
(Fig. \ref{fig:vel_r_phase}). The massive, rotationally-supported disk is decoupled from the feedback loop; the radially-dominant infalling cold gas 
is what closes the feedback cycle. The cold torus rotates at the 
local circular speed (200-300 km s$^{-1}$). The infalling cold gas can be fast ($\lesssim 400$ km s$^{-1}$), but the uplifted cold gas from the rotating torus 
can sometimes reach speeds larger than 1000 km s$^{-1}$ as it is accelerated by the fast jet (Fig. \ref{fig:velpdf}). The mass of the radially-dominant infalling 
cold gas is about a factor of two times the outflowing cold gas. The massive cold torus is expected to be substantially depleted by star formation, which we do 
not take into account in our simulations.

\item The minimum in the ratio of the cooling time and the free fall time (min[\tr]) seems better than the core entropy ($K_0$) for characterizing the cool 
cores. First of all it is a dimensionless parameter which applies for all halo masses (\citealt{voi14b}), and secondly
it is not sensitive to strong cooling or heating in the very center (unlike $K_0$). The entropy and \tr~panels in Figure \ref{fig:entropy} show that a constant \tr~`core', 
which corresponds to a double power law for the entropy profile (with a slow increase with radius in the core; as argued in \citealt{pan14}), is a better 
approximation to clusters in the very cool state with min(\tr)$ \lesssim 5$. 

\end{itemize}

\acknowledgments
This work is partly supported by the DST-India grant no. Sr/S2/HEP-048/2012 and an India-Israel joint research grant (6-10/2014[IC]). 
DP is supported by a CSIR grant (09/079[2599]/2013-EMR-I). AB acknowledges funding from NSERC Canada through the Discovery Grant program.


\begin{thebibliography}{}
\bibitem[Babul et al. (2002)]{bab02} Babul, A., Balogh, M. L., Lewis, G. F., \& Poole, G. B. 2002, MNRAS, 330, 329 
\bibitem[Babul et al. (2013)]{bab13} Babul, A., Sharma, P., \& Reynolds, C. S. 2013, \apj, 768, 11
\bibitem[Balogh, Babul, \& Patton (1999)]{bal99} Balogh, M. L., Babul, A., \& Patton, D. R. 1999, \mnras, 307, 463
\bibitem[Banerjee \& Sharma (2014)]{ban14} Banerjee, N. \& Sharma, P. 2014, \mnras, 443, 687
\bibitem[Benson \& Babul (2009)]{ben09} Benson, A. J. \& Babul, A. 2009, \mnras, 397, 1302
\bibitem[Bildfell et al. (2008)]{bil08} Bildfell, C., Hoekstra, H., Babul, A., \& Mahdavi, A. 2008, \mnras, 389, 1637
\bibitem[Binney \& Tabor (1995)]{bin95} Binney, J. \& Tabor, G. 1995, \mnras, 276, 663
\bibitem[Birnboim \& Dekel (2003)]{bir03} Birnboim, Y. \& Dekel, A. 2003, \mnras, 345, 349
\bibitem[B\^irzan et al. (2004)]{bir04} B\^irzan, L., Rafferty, D. A., McNamara, B. R., Wise, M. W., \& Nulsen, P. E. J. 2004, \apj, 607, 800
\bibitem[B\"ohringer et al. (2002)]{boh02} B\"ohringer, H., Matsushita, K., Churazov, E., Ikebe, Y., \& Chen, Y. 2002, A\&A, 382, 804
\bibitem[Cattaneo \& Teyssier (2007)]{cat07} Cattaneo A. \& Teyssier R., 2007, MNRAS, 376,1547
\bibitem[Cavagnolo et al. (2008)]{cav08} Cavagnolo K. W., Donahue M., Voit M. \& Sun M., 2008, ApJ, 683,107
\bibitem[Cavagnolo et al. (2009)]{cav09} Cavagnolo, K. W., Donahue, M., Voit, G. M., \& Sun, M. 2009, \apjs, 182, 12  
\bibitem[Cielo et al. (2014)]{cie14} Cielo, S., Antonuccio-Delogu, V., Macci\'o, A. V., Romeo, A. D., \& Silk, J. 2014, \mnras, 439, 2903
\bibitem[Ciotti \& Ostriker (2001)]{cio01} Ciotti, L., \& Ostriker J. 2001, ApJ, 551, 131
\bibitem[Crawford et al. (1999)]{cra99} Crawford, C. S., Allen, S. W., Ebeling, H., Edge, A. C., \& Fabian, A. C. 1999, \mnras, 306, 857
\bibitem[David et al. (2014)]{dav14} David, L. P. et al. 2014, \apj, 792, 94
\bibitem[Dekel \& Birnboim (2008)]{dek08} Dekel, A. \& Birnboim, Y. 2008, \mnras, 383, 119
\bibitem[De Lucia \& Blaizot (2007)]{del07} De Lucia, G. \& Blaizot, J. 2007, \mnras, 375, 2
\bibitem[Dennis \& Chandran (2005)]{den05} Dennis, T. J. \& Chandran, B. D. G. 2006, \apj, 622, 205
\bibitem[Donahue et al. (2000)]{don00} Donahue, M. et al. 2000, \apj, 545, 670
\bibitem[Dubois et al. (2010)]{dub10} Dubois, Y., Devriendt, A., Slyz, A., \& Teyssier, R. 2010, \mnras, 409, 985 
\bibitem[Edge (2001)]{edg01} Edge, A. C. 2001, \mnras, 328, 762
\bibitem[Edge et al. (2010)]{edg10} Edge, A. C., Oonk, J. B. R., Mittal, R. et al. 2010, A\&A, 518, L46
\bibitem[Fabian (1994)]{fab94} Fabian A. C., 1994, ARAA, 32, 277F
\bibitem[Fabian et al. (2003)]{fab03} Fabian, A. C. et al. 2003, \mnras, 344, L43
\bibitem[Gaspari et al. (2012)]{gas12} Gaspari, M., Ruszkowski, M., \& Sharma, P., 2012, ApJ, 746, 94
\bibitem[Gaspari et al. (2013)]{gas13} Gaspari M., Ruszkowski M., \& Oh S. P., 2013, MNRAS, 432, 3401
\bibitem[Gaspari et al. (2014)]{gas14} Gaspari, M., Ruszkowski, M., Oh, S. P., Brighenti, F., \& Temi, P. 2014, arXiv:1407.7531
\bibitem[Hayes et al. (2006)]{hay06} Hayes, J. C.  et al. 2006, \apjs, 165, 188
\bibitem[Heinz et al. (2006)]{hei06} Heinz, S., Br\"uggen, M., Young, A., \& Levesque, E. 2006, \mnras, 373, L65
\bibitem[Hicks, Mushotzky, \& Donahue (2010)]{hic10} Hicks, A. K., Mushotzky, R., \& Donahue, M. 2010, \apj, 719, 1844
\bibitem[Hobbs et al. (2011)]{hob11} Hobbs, A., Nayakshin, S., Power, C., \& King, A. 2011, \mnras, 413, 2633
\bibitem[Kaiser (1986)]{kai86} Kaiser, N. 1986, \mnras 222, 323
\bibitem[Kaiser (1991)]{kai91} Kaiser, N. 1991, \apj, 383, 104
\bibitem[Lewis et al. (2000)]{lew00} Lewis, G. F., Babul, A., Katz, N., Quinn, T., Hernquist, L., \& Weinberg, D. H. 2000, \apj, 536, 623
\bibitem[Li \& Bryan (2014a)]{li14} Li, Y. \& Bryan, G. L. 2014a, \apj, 789, 153
\bibitem[Li \& Bryan (2014b)]{li14b} Li, Y. \& Bryan, G. L. 2014b, \apj, 789, 54
\bibitem[Li et al. (2015)]{li15} Li, Y., Bryan, G., Ruszkowski, M., Voit, G. M., O'Shea, B. W., \& Donahue, M. 2015, arXiv:1503.02660  
\bibitem[Lim et al. (2008)]{lim08} Lim, J., Ao, Y., \& Dinh-V-Trung 2008, \apj, 672, 252
\bibitem[Loewenstein et al. (2001)]{loe01} Loewenstein, M., Mushotzky, R. F., Angelini, L., Arnaud, K. A., \& Quataert, E. 2001, \apj, 555, L21 
\bibitem[McCarthy et al. (2008)]{mcc08} McCarthy, I. G., Babul, A., Bower, R. G., \& Balogh, M. L. 2008, \mnras, 386, 1309
\bibitem[McCourt et al. (2012)]{mcc12} McCourt, M., Sharma, P., Quataert, E. \& Parrish, I. J. 2012, \mnras, 419. 3319
\bibitem[McDonald et al. (2010)]{mcd10} McDonald, M., Veilleux, S., Rupke, D. S. N., \& Mushotzky, R. 2010, \apj, 721, 1262
\bibitem[McDonald et al. (2011a)]{mcd11} McDonald M., Veilleux S., \& Mushotzky R., 2011, ApJ, 731,33
\bibitem[McDonald et al. (2011b)]{mcd11b} McConald, M., Veilleux S., Rupke, D. S. N., Mushotzky R., \& Reynolds, C. S. 2011, \apj, 734, 95
\bibitem[McNamara \& Nulsen, 2007]{mcn07} McNamara B. R., \& Nulsen P. E. J., 2007, ARA\&A, 45, 117
\bibitem[McNamara, Rohanizadegan, \& Nulsen (2011)]{mcn11} McNamara, B. R., Rohanizadegan, M., \& Nulsen, P. E. J. 2011, \apj, 727, 39
\bibitem[McNamara et al. (2014)]{mcn14} McNamara, B. R. et al. 2014, \apj, 785, 44
\bibitem[Merloni et al.(2003)]{mer03} Merloni, A. Heinz, S., \& Di Matteo, T. 2003, \mnras, 345, 1057 
\bibitem[Mignone et al. (2013)]{mig13} Mignone, A., Striani, E., Tavani, M., \& Ferrari, A. 2013, \mnras, 436, 1102
\bibitem[Mittal et al. (2009)]{mit09} Mittal, R., Hudson, D. S., Reiprich, T. H., \& Clarke, T. 2009, A\&A, 501, 835
\bibitem[Navarro et al., 1996]{nav96} Navarro J. F., Frenk C. S., \& White S. D., 1996, ApJ, 462, 563
\bibitem[Narayan \& Yi (1995)]{nar95} Narayan, R. \& Yi, I. 1995, \apj, 444, 231
\bibitem[O'Dea et al. (2008)]{ode08} O'Dea, C. P. et al. 2008, \apj, 681, 1035
\bibitem[Omma et al. (2004)]{omm04} Omma H., Binney J., Bryan G., \& Slyz A., 2004, MNRAS, 348, 1105
\bibitem[Omma \& Binney (2004)]{omm04b} Omma H., \& Binney J., 2004, MNRAS, 350, L13
\bibitem[O'Sullivan et al. (2015)]{osu15} O'Sullivan, E., Combes, F., Hamer, S., Salom\'e, P., Babul, A., \& Raychaudhury, S. 2015, A\&A, 573, A111 
\bibitem[Panagoulia, Fabian, \& Sanders (2014)]{pan14} Panagoulia, E. K., Fabian, A. C., \& Sanders, J. S. 2014, \mnras, 438, 2341
\bibitem[Peterson et al. (2003)]{pet03} Peterson, J. R., Kahn, S. M., Faerels, F. B. S. et al. 2003, \apj, 590, 207
\bibitem[Pizzolato \& Soker (2005)]{piz05} Pizzolato, F. \& Soker, N. 2005, \apj, 632, 821
\bibitem[Pizzolato \& Soker (2010)]{piz10}  Pizzolato, F. \& Soker, N. 2010, \mnras, 408, 961
\bibitem[Ponman et al. (1999)]{pon99} Ponman, T. J., Cannon, D. B., \& Navarro, J. F. 1999, Nature, 397, 135
\bibitem[Poole et al. (2008)]{poo08} Poole, G. B., Babul, A., McCarthy, I. G., Sanderson, A. J. R., \& Fardal, M. A. 2008, \mnras, 391, 1163
\bibitem[Pope et al. (2010)]{pop10} Pope, E. C. D., Babul, A., Pavlovski, G., Bower, R. G., Dotter, A. 2010, \mnras, 406, 2023
\bibitem[Pratt et al. (2009)]{pra09} Pratt, G. W., Croston, J. H., Arnaud, M., \& B\"ohringer, H. 2009, A\&A, 498, 361
\bibitem[Rafferty, McNamara, \& Nulsen (2008)]{raf08} Rafferty, D. A., McNamara, B. R., \& Nulsen, P. E. J. 2008, \apj, 687, 899
\bibitem[Rawle et al. (2012)]{raw12} Rawle, T. D., Edge, A. C., Egami, E. et al. 2012, 747, 29
\bibitem[Revaz, Combes, \& Salom\'e (2008)]{rev08} Revaz, Y., Combes, F., \& Salom\'e 2008, A\&A, 477, L33
\bibitem[Robertson et al. (2010)]{rob10} Robertson, B. E., Kravtsov, A. V., Gnedin, N. Y., Abel, T., \& Rudd, D. H. 2010, \mnras, 401, 2463 
\bibitem[Russell et al. (2014)]{rus14} Russell, H. R., McNamara, B. R., Edge, A. C. et al. 2014, \apj, 784, 78
\bibitem[Salom\'e et al. (2006)]{sal06} Salom\'e, Combes, F., Edge, A. C., et al. 2006, A\&A, 454, 437
\bibitem[Saro et al. (2006)]{sar06} Saro, A., Borgani, S., Tornatore, L., Dolag, K., Murante, G., Biviano, A., Calura, F., \& Charlot, S. 2006, \mnras, 373, 397
\bibitem[Sharma, Parrish, \& Quataert (2010)]{sha10} Sharma, P., Parrish, I. J., \& Quataert, E. 2010, \apj, 720, 652
\bibitem[Sharma et al. (2012a)]{sha12} Sharma P., McCourt M., Quataert E., \& Parrish I. J., 2012, MNRAS, 420, 3174
\bibitem[Sharma et al. (2012b)]{sha12b} Sharma, P., McCourt, M., Parrish, I. J., \& Quataert, E. 2012, \mnras, 427, 1219
\bibitem[Singh \& Sharma (2015)]{sin15} Singh, A. \& Sharma, P. 2015, \mnras, 446, 1895
\bibitem[Soker et al., 2001]{sok01} Soker N., White III R. E., David L. P., \& McNamara B. R., 2001, ApJ, 549, 832
\bibitem[Sternberg, Pizzolato, \& Soker (2007)]{ste07} Sternberg, A., Pizzolato, F., \& Soker, N. 2007, \apj, 656, L5
\bibitem[Stone \& Norman (1992)]{sto92} Stone, J. M. \& Norman, M. L. 1992, \apjs, 80, 753
\bibitem[Sun (2009)]{sun09} Sun, M. 2009, \apj, 704, 1586
\bibitem[Tombesi et al. (2010)]{tom10} Tombesi, F. et al. 2010, \apj, 719, 700
\bibitem[Tremblay et al. (2012)]{tre12} Tremblay, G. R., O'Dea, C. P., Baum, S. A. et al. 2012, arXiv1205.2373
\bibitem[Vernaleo \& Reynolds, 2006]{ver06} Vernaleo J. C., \& Reynolds C. S., 2006, ApJ, 645, 83
\bibitem[Voigt \& Fabian, 2004]{voi04} Voigt L. M., \& Fabian A. C., 2004, MNRAS, 347, 1130
\bibitem[Voit (2005)]{voi05} Voit, G. M. 2005, Rev. Mod. Phys., 77, 207
\bibitem[Voit (2011)]{voi11} Voit, G. M. 2011, \apj, 740, 28
\bibitem[Voit \& Donahue (2014)]{voi14} Voit, G. M. \& Donahue, M. 2014, arXiv1409.1601
\bibitem[Voit  et al. (2015a)]{voi14b} Voit, G. M., Donahue, M., Bryan, G. L., \& McDonald, M. 2015, 519, 203
\bibitem[Voit et al. (2015b)]{voi15} Voit, G. M., Donahue, M., O'Shea, B. W., Brian, G. L., Sun, M., \& Werner, N. 2015, arXiv:1503.02104
\bibitem[Wagh et al. (2014)]{wag14} Wagh, B., Sharma, P., \& McCourt, M. 2014, \mnras, 439, 2822 
\bibitem[Werk et al. (2014)]{werk14} Werk, J. K. et al. 2014, \apj, 792, 8
\bibitem[Werner et al. (2014)]{wer14} Werner, N. et al. 2014, \mnras, 439, 2291
\end{thebibliography}
\end{document}